# Early Atomic Models – From Mechanical to Quantum (1904-1913)

Charles Baily
Department of Physics
University of Colorado
Boulder, CO 80309-0390, USA

## Abstract

A complete history of early atomic models would fill volumes, but a reasonably coherent tale of the path from mechanical atoms to the quantum can be told by focusing on the relevant work of three great contributors to atomic physics, in the critically important years between 1904 and 1913: J. J. Thomson, Ernest Rutherford and Niels Bohr. We first examine the origins of Thomson's mechanical atomic models, from his ethereal vortex atoms in the early 1880's, to the myriad "corpuscular" atoms he proposed following the discovery of the electron in 1897. Beyond qualitative predictions for the periodicity of the elements, the application of Thomson's atoms to problems in scattering and absorption led to quantitative predictions that were confirmed by experiments with high-velocity electrons traversing thin sheets of metal. Still, the much more massive and energetic $\alpha$-particles being studied by Rutherford were better suited for exploring the interior of the atom, and careful measurements on the angular dependence of their scattering eventually allowed him to infer the existence of an atomic nucleus. Niels Bohr was particularly troubled by the radiative instability inherent to any mechanical atom, and succeeded in 1913 where others had failed in the prediction of emission spectra, by making two bold hypotheses that were in contradiction to the laws of classical physics, but necessary in order to account for experimental facts.

## Contents

# 1    Introduction

Tremendous strides were made in the nascent field of atomic physics during the relatively short time between the discovery of the electron in 1897, and the birth of the quantum atom in 1913. Beginning with almost no understanding of atoms other than their chemical and spectral properties, physicists were handed important clues to their internal structure with the discovery of spontaneous radiation, first identified by Becquerel in 1896 by its ability to produce a photographic effect. The very existence of atomic radiation strongly suggested that atoms were not indivisible after all, and when Joseph John Thomson (1856–1940) announced in 1897 that *cathode rays* were actually comprised of negatively charged particles, he was already convinced that these "corpuscles" must be fundamental constituents of matter.

It had been known for some time from Maxwell's theory that accelerated charges were responsible for the production of electromagnetic waves, and there seemed to be no doubt that atomic spectra must be due to the motion of these discrete charges within the atom. The corpuscular model proposed by Thomson in 1904 was poorly suited for predicting spectral lines, but he demonstrated that his mechanical atom, a uniformly charged sphere embedded with rotating rings of electrons, had an amazing explanatory power for the observed periodicity in the elements. Thomson later applied modified versions of this model to a variety of physical phenomena, such as the dispersion of light by dilute gases, and developed methods for estimating the actual number of electrons in an atom, which he concluded must be roughly equal to its atomic weight, and not the "thousands" suggested by the small mass-to-charge ratio of an electron. By 1910, experiments had confirmed many of his model's predictions for the absorption and scattering of electrons in thin materials.

Indeed, the radiation spontaneously produced by atoms eventually became the very tool used by physicists to probe their internal workings. In 1898, Ernest Rutherford (1871–1937) had been able to distinguish two types of atomic radiation (α & β) by the difference in their ability to penetrate matter. In almost complete ignorance of their basic nature, Rutherford gradually increased the complexity of the experimental questions he posed.  Are the α-rays deflected by a magnetic field?  Are the α-particles positively or negatively charged?  What is the magnitude of their charge?  How much kinetic energy do they lose when passing through thin sheets of aluminum?  The scattering of α-particles by matter was significantly less pronounced than for β-particles, but ultimately noticeable, and Rutherford's ongoing experiments inspired a series of careful measurements by Hans Geiger and Ernest Marsden on the degree of scattering and reflection caused by various types and thicknesses of metal.  Rutherford used this data in 1911 to show that large-angle scattering could be explained in terms of *single* encounters with a massive nuclear core, but not by *multiple* encounters with a positively charged sphere of atomic dimensions, as was Thomson's view.  The formula derived by Thomson assuming small-angle *compound* scattering would only generate appropriate numbers if the radius of the sphere were reduced by several orders of magnitude.

Danish physicist Niels Bohr (1885–1962), who spent the better part of a post-doctoral year with Thomson in Cambridge before being invited to work with Rutherford at the University of Manchester in 1912, was deeply troubled by the use of mechanical models to describe atomic spectra.  This even despite recent success by J. W. Nicholson at matching the orbital frequencies of his mechanical (and nuclear) model with specific lines in the solar corona, by restricting changes in the angular momenta of his electron rings to whole units of Planck's constant.  Bohr's profound insight was that the discrete nature of line-spectra could not be explained in terms of the periodic motion of atomic charges, for this would require them to orbit at constant frequencies for a finite amount of time.  If the laws of electrodynamics were universally valid, their immediate loss of kinetic energy through the radiation they produced would actually predict a continuous emission spectrum.  The quantum rules he invented to account for this discrepancy had no basis in the well-established laws of physics, but found some justification, Bohr claimed, in their correspondence with classical expectations in the regime of large quantum numbers.  Nevertheless, the unprecedented success of his quantum model at predicting the visible spectra of hydrogen and other single-electron atoms (in terms of fundamental constants, no less) eventually led to its widespread adoption, sowing the seeds of the quantum revolution. Today, mechanical atoms are little more than historical curiosities.

When exploring the early development of mechanical models of the atom, one is naturally interested in learning what originally inspired their salient features, and what mathematical techniques were employed to deduce their properties based on those features. Thomson's first foray into atomic modeling came in his 1882 Adams Prize-winning essay on the dynamics of vortices in an ideal fluid, wherein he articulated a sophisticated theory of atoms as stable vortices in the electromagnetic ether. The concept of ethereal vortex atoms had been proposed in 1867 by Sir William Thomson (later, Lord Kelvin), and he was indebted to Helmholtz[1] for the mathematics he used to describe them. Interestingly, even authoritative histories typically fail to mention the remarkable similarities between Thomson's investigation into the stability of rotating vortex rings and the methods used by James Clerk Maxwell (1831–1879) in his treatise on the dynamics of Saturn's rings (also awarded the Adams Prize in 1857). The omission of this one fact seems entirely arbitrary, considering the way historians of atomic modeling generally acknowledge the pervasive influence of Maxwell in so many aspects of modern physics.

For example, Kragh mentions Maxwell only twice in the introductory chapter (on pre-quantum atoms) of his recent book about the Bohr model;[2] first, for his written praise of Kelvin's vortex models in the 1875 *Encyclopaedia Britannica*; [Kragh, p. 6] second, and most naturally for this context, in connection with the "Saturnian" atomic model proposed by Hantaro Nagaoka in 1904, which was based on Maxwell's calculations. [Kragh, p. 23] In a collection of historical essays on atomic structure,[3] Heilbron calls attention to the influence on an entire era of Victorian physics of Maxwell's predilection for mechanical analogies. A scientific biography by Davis and Falconer[4] describes J. J. Thomson's youthful devotion to Maxwellian electrodynamics and the mechanical ether.

---

[1] Thomson, W., p. 15. An English translation by Tait had recently appeared [*Phil. Mag.* **33:** 485-512] of Helmholtz, H. 1858. Ueber Integrale der hydrodynamischen Gleichungen, welche den Wirbelbewegungen entsprechen. *Journal für die reine und angewandte Mathematik* **55:** 25-55. William Thomson also makes mention of papers from Rankine (1849-50) on "Molecular Vortices".

[2] Kragh, H. 2012. *Niels Bohr and the Quantum Atom*, Oxford University Press, Oxford.

[3] Heilbron, J. L. 1981. *Historical Studies in the Theory of Atomic Structure*, Arno Press, New York.

Moreover, Maxwell's legacy as the visionary founder of the Cavendish Laboratory in Cambridge is one of the many threads that bind the three main actors in the story that follows. This is the place where Thomson was appointed as director in 1884, where Rutherford worked as Thomson's first research student from 1895 to 1898, and where Bohr stopped over in 1911 before moving on to join Rutherford at the University of Manchester. There is an old adage that most lines of research in modern physics, when traced back far enough, will eventually lead to James Clerk Maxwell, and this is no less true when delving into the origins of mechanical models of the atom.

---

# 2 The mechanical atoms of J. J. Thomson

## 2.1 Rings of Saturn and ethereal vortices

The introduction to Maxwell's 1857 essay, "On the Stability of the Motion of Saturn's Rings," contains a concise statement of the central theme of his analysis, but also that of an entire research program yet to come on mechanical models of the atom.

> "Having found a particular solution of the equations of motion of any material system, to determine whether a slight disturbance of the motion indicated by the solution would cause a small periodic variation, or a total derangement of the motion." [Maxwell 1859, p. 5]

The prize committee for that year had asked if the long-term stability of Saturn's rings could be explained on dynamical principles, under the assumption they were either solid, liquid, or made up from particulate matter. In answer to this challenge, Maxwell simultaneously brought to bear a number of mathematical techniques (in particular, Fourier analysis and Lagrangian mechanics) to first show that a uniformly solid ring would be *dynamically unstable*,[5] and that a liquid ring must ultimately break apart into disconnected droplets.

The remaining possibility was for the rings to be comprised of independent particles (whether solid or liquid), each moving under the gravitational influence of the central mass, as well as that of all the other orbiting particles. Maxwell approximated a single planetary ring as a collection of point masses distributed at equal-angle intervals around a circle, derived the equations of motion for two masses in stable orbit, then stepwise let the number of satellites grow until arbitrarily large. Upon determining the conditions for steady-state motion, he considered the effect of any small deviation of the particles from their orbits, in the

---

[5] There would be an insufficient restoring force if the centers of mass for the uniform ring and the planet ever deviated from equilibrium, eventually leading the ring and planet to crash into each other. The only exception was most unlikely: an otherwise uniform ring would require an additional point mass located at its outer edge, equal to 0.82 of the mass of the total ring. [Maxwell 1859, p. 55]

radial and tangential directions, as well as perpendicular to the plane of rotation. If the motion of the particles were to be permanent, the solutions to these sets of differential equations would have to all be sinusoidal (the particles would oscillate about their equilibrium positions); exponential solutions corresponded to the ring breaking apart. This placed basic constraints on the properties of the system – most notably, the mass of the central body would have to be substantially larger than the sum of the orbiting masses. [Maxwell 1859, p. 25]

The influence of Maxwell is readily apparent in Thomson's own 1882 Adams Prize essay. This time, the committee had called for a "general investigation of the action upon each other of two closed vortices in a perfect incompressible fluid," a topic that was of current interest to mathematical physicists, particularly since Maxwell had described Faraday's *lines of force* in terms of vortices in an electromagnetic medium.[6] The fundamental simplicity of Kelvin's more recent ideas appealed to J. J. Thomson, that all the various properties of materials might be explained in terms of the dynamics of ethereal vortices.

> "The equations which determined this motion were known from the laws of hydrodynamics, so that if the theory were true the solution of the problem of the universe would be reduced to the solution of certain differential equations, and would be entirely a matter of developing mathematical methods powerful enough to deal with what would no doubt be very complex distributions of vortex motion in the fluid. [...] The investigation ... involved long and complicated mathematical analysis and took a long time. It yielded, however, some interesting results and ideas which I afterwards found valuable in connection with the theory of the structure of the atom..." [Thomson 1936, p. 94-5]

The first two parts of Thomson's essay dealt with the mathematics describing a single vortex in an ideal fluid, and then two vortices separated by a distance large compared to their radii, which alone might have been sufficient to satisfy the

---

[6] Maxwell 1861, p. 165.

requirements of the committee. The third and fourth parts developed a comprehensive theory of vortex atoms, where Thomson first considered the interaction of two vortices in close proximity, finding that stable uniform motion was possible when they were of equal strength and revolved about a common center. For three or more identical vortices, all arranged in a plane at regular intervals around the circumference of a circle (he made clear this was not the most general configuration), any small disturbance would cause them to execute stable oscillations about their equilibrium positions, but only if there were less than seven of them. Thomson calculated the periods of these oscillations for each case, and demonstrated that the system would be dynamically unstable if there were more than six vortices in a ring. [Thomson 1883a, p. 107-8]

Beyond mathematical expediency, Thomson's choice to work with symmetrically distributed vortices could be physically motivated with an analogy to the static arrangements of floating magnets under the influence of a central force, as reported by Alfred Mayer in 1878. Mayer had stuck magnetized needles into pieces of cork with the south ends pointing upward, so that they were all mutually repelled, but each attracted to the north end of a bar magnet placed at the center. This combination of push and pull caused the floating needles to rearrange themselves into regular polygons, just like Thomson's vortices. [Fig. 1] With $n = 5$, *two* static arrangements were possible: a pentagon [5a] and a square with a needle at the center [5b]; but only the pentagon was stable against external perturbations. Static rings of six, seven and more were possible with additional interior needles, which themselves settled into stable patterns as their numbers grew (i.e., the configuration for 5a could form the core of $n = 14$, whereas 5b could not). [Mayer, p. 251] Mayer himself thought these simple rules might provide insight into various molecular properties (such as *allotropy* and *isomerism*), but it was the future Lord Kelvin who first made the connection between the static configurations of the magnets and those for vortex atoms in a state of mechanical equilibrium.[7]

---

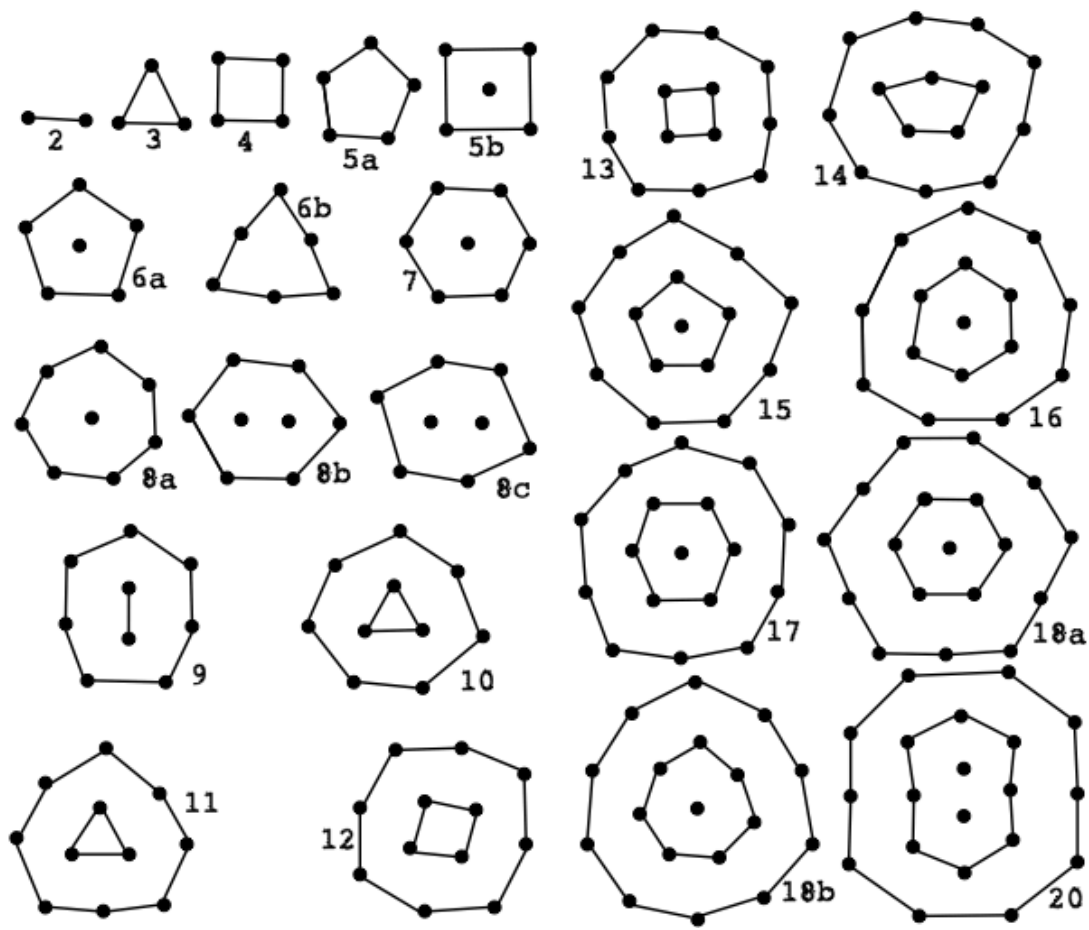

**Figure 1.** Static arrangements for $n = 2 - 20$ floating magnetic needles (all with their south poles oriented vertical), under the attractive influence of a central bar magnet. If multiple static configurations were possible for a given $n$, only those denoted by an "a" were stable against perturbations. [From Mayer, p. 248-9]

J. J. Thomson also used his vortex model to explain the chemical combination of elements, by associating the number of *primary* vortices in an atom with its valency (each primary vortex could just as well be replaced by a stable subsystem of *secondary* vortices, if the net strength of the subsystem equaled the strength of the other primaries). Thus, an atom containing two vortices (a *dyad*) might combine with two atoms having each a single vortex (a *monad*), to form a stable atom with two primaries (each a subsystem of two secondary vortices). The myriad combinations of monads, dyads, triads and so on were innumerable, but constrained by a requirement that the vortices divide themselves into primaries of equal strength, and that there not be more than six of them on a ring. This latter finding also agreed with chemical facts, Thomson claimed, in that there were no known gaseous compounds of one element combined with more than six atoms of another. [Thomson 1883a, p. 120-21]

By the time of his appointment as Cavendish Professor in 1884, Thomson had already turned to the study of electrical discharges in gases, which he thought could also be explained in terms of vortices and ethereal strains. In 1883, he described an externally applied electric field as a velocity gradient in the ether directed along the *lines of force*. The overall effect of this non-uniform velocity field would be to separate paired vortices, with the resulting relaxation in the ether corresponding to electricity passing through the gas (much like Maxwell's notion of an ethereal *displacement current*). [Thomson 1883b, p. 428-29]

Thomson's early vortex theory of electrical discharge was based on the concept of energy dissipation, but at the time there was mounting experimental evidence for the existence of discrete charges, and he eventually began to think about the electrical breakdown of a gas as the separation and transfer of charge.[8] In the ensuing years prior to 1897, his continuing investigations ultimately led to a series of experiments on the rays emitted by the cathode during a spark. The deflection of their trajectories by a magnetic field showed that they were not like X-rays, but rather high velocity, negatively charged 'corpuscles' of matter. [Thomson, 1897]

## 2.2    A corpuscular theory of matter

"After long consideration of the experiments it seemed to me that there was no escape from the following conclusions: (1) That atoms are not indivisible, for negatively electrified particles can be torn from them by the action of electrical forces, impact of rapidly moving atoms, ultraviolet light or heat.

---

[8] His first conceptions were in analogy with electrolysis. [Davis and Falconer, p. 77-8; see also p. 77-138 for details on the evolution in Thomson's thinking about gaseous discharge leading up to his discovery of the electron.]

(2) That these particles are all of the same mass, and carry the same charge of negative electricity from whatever kind of atom they may be derived, and are a constituent of all atoms. (3) That the mass of these particles is less than one-thousandth part of the mass of an atom of hydrogen." [Thomson 1936, p. 338-9]

Convinced that these corpuscles were universal components of matter, Thomson decided to relate what he already knew about vortex atoms to bound systems of discrete charges. He again drew attention to Mayer's magnets, asserting that the periodicity in the configurations of concentric rings suggested an analogy with the periodicity in the chemical and spectral properties of the elements; but in 1897 he had yet to propose anything more than a qualitative picture of corpuscular atoms. Being all of the same charge, a dynamically stable system of corpuscles would be impossible without some kind of compensating positive charge to hold them all together (like Mayer's bar magnet to the needles).[9] Unable to explain the electrical neutrality of an atom as a manifestation of its negatively charged constituents, Thomson was left with no choice but to concoct a hypothetical smear of positive atomic charge, which he likened (crudely) to "a liquid with a certain amount of cohesion, enough to keep it from flying to bits under its own repulsion."[10] Thomson never believed in the literal truth of his mechanical analogies, and preferred to de-emphasize any physical interpretation of this "positive electrification." [Davis and Falconer, p. 195]

Regardless of whether the positive charge were supposed to have any substance (or even reality), the equations he derived in 1904 described the periodic motion of corpuscles in a positively charged region of space, and contained no viscous damping terms, meaning the electrons were free to move about his atom without resistance. Thomson's dynamic model could therefore be viewed as a

---

[9] Positively charged subatomic particles had been proposed but never observed. Goldstein had discovered positively charged rays during cathode experiments reported in 1886, but they were only known at the time to have a charge-to-mass ratio similar to ionized hydrogen. [Davis and Falconer, p. 199] It was not until after the concept of *isotopy* was developed that physicists could identify these "positive rays" as singly-ionized *tritium* (an isotope of hydrogen, with one proton and two neutrons), and not triatomic hydrogen (as proposed by Thomson in 1913). [Kragh, p. 96-100]

[10] Letter to Sir Oliver Lodge from 11 April 1904. [Quoted in Davis and Falconer, p. 195-96]

generalization of the atom expounded by Kelvin in an article from 1902, where he revisited the "one fluid theory of electricity" of Franz Aepinus (1724–1802). As described by Kelvin, Aepinus proposed the existence of a negatively charged electric fluid that permeated all of space, and flowed freely "among the atoms of ponderable matter." Negative (and positive) charge distributions could thus be viewed as regions of space with an excess (or deficiency) of this ideal fluid. Kelvin's proposal was that such a fluid could actually be made up from a multitude of tiny, negatively charged "atoms of electricity"[11] that moved without resistance inside of positively charged, spherical atoms. [Kelvin, p. 257]

Kelvin's aim was to deduce the spatial configurations of negative point charges[12] in a state of static equilibrium within a positively charged atom. Two of his explicit assumptions[13] were that: (A) positively charged atoms repel each other according to an inverse square law (in accordance with the discoveries of Cavendish and Coulomb); and (B) negatively charged particles within an atom are attracted to its center by a force that is directly proportional to their radial distance. [Kelvin, p. 258] A straightforward application of Gauss' law would show that the interior field of this atom could only be radially outward and linearly increasing if the positive charge were spherically symmetric and uniformly distributed.

Kelvin's model was static, but Thomson was interested in the dynamic stability of atomic corpuscles executing orbital motion, which he physically motivated by suggesting a connection with radioactivity:

> "...suppose the atoms of a substance, like the atoms of radio-active substances, were continually emitting corpuscles; the velocity of the corpuscles under consideration being, however, insufficient to carry them clear of the atom, so that the corpuscles describe orbits round the centre of the atom..." [Thomson 1903, p. 689]

---

[11] Hence the title of his paper, "Aepinus Atomized."

[12] Kelvin asserted that these negatively charged *electrions* "no doubt occupy finite spaces, although at present we are dealing with them as if they were mere mathematical points…" [Kelvin, p. 258]

[13] "As a tentative hypothesis, I assume *for simplicity* that…" [Kelvin, p. 258; emphasis added.]

In 1903, Thomson had rigorously investigated the electromagnetic fields produced by a steadily rotating system of charges, arranged in a plane at equal intervals around a ring (but in the absence of any positive charge). This was explicitly done in anticipation of future work, and his results would have important implications for the radiative stability of his atoms.

The dynamical atom initially proposed by Thomson in March 1904 consisted of a uniform sphere of positive charge and a single ring of negatively charged corpuscles[14] (arranged symmetrically, as before). [Fig. 2] The large charge-to-mass ratio for electrons, as well as the numerous spectral lines showing the Zeeman effect,[15] had certainly implied the presence of thousands of atomic corpuscles,[16] but none of the atoms in this paper had more than a few hundred, and the majority of his calculations were devoted to rings with less than seven charges. Similar to Mayer's magnets and the particulates of Saturn's rings (but with a different central force law), the motion of each charge was governed only by its electrostatic attraction toward the center of the sphere, and by the repulsion from each of the other negative charges.

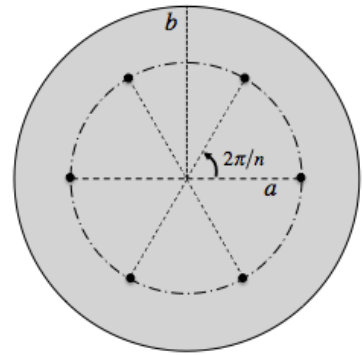

**Figure 2.** Diagram of Thomson's 1904 atomic model.[17] A uniform sphere of positive charge (shaded region, of radius $b$) contains $n$ negative point charges arranged at equal intervals around a circle (of radius $a$). The ratio $a/b = 0.6726$ for a static ring of $n = 6$ charges (see below).

---

[14] Thomson continued to refer to negative atomic charges as "corpuscles" throughout the article, and for many years after. George Stoney claimed in 1894 to have coined the term *electron*; Kelvin used the term *electrions* for his 'atoms of electricity'. Rutherford's 1911 nuclear model was agnostic on the actual form and distribution of the negative charge. Nagaoka (1904), Haas (1910), Nicholson (1912) and Bohr (1913) all identified their negative atomic charges as electrons.

[15] The splitting of spectral lines emitted by excited atoms in a magnetic field. It was assumed that there was a one-to-one correspondence between each Zeeman line and an orbiting electron.

[16] If the positive charge distribution had no inertial mass, there would be no constraint on the amount of positive charge available to neutralize an atom filled with thousands of electrons.

[17] He did not reproduce any illustrations of his model, perhaps because it was so simple that a picture would be superfluous. Except for the positive charge distribution, it is clearly described in the title alone: "On the Structure of the Atom: an Investigation of the Stability and Periods of Oscillation of a number of Corpuscles arranged at equal intervals around the Circumference of a Circle; with Application of the results to the Theory of Atomic Structure."

After briefly introducing his model, Thomson quickly set about analyzing a system of $n$ particles (mass $m$ and charge $-e$), arranged on a ring of radius $a$, centered within a positive sphere of radius $b$. For *static* equilibrium, the radially attractive force on each charge (equal to $ne^2a/b^3$ for an electrically neutral atom) must balance the net repulsive force of the other $n-1$ charges. The azimuthal component of this repulsive force would be zero due to symmetry, and the radial component $F_r$ given by:[18]

$$F_r = \sum_{k=1}^{n-1} \left|\vec{\mathbf{F}}_k\right| \sin\left(k\pi/n\right) = \sum_{k=1}^{n-1} \frac{e^2 \sin\left(k\pi/n\right)}{\left(a - a\cos\left(2k\pi/n\right)\right)^2 + a^2 \sin^2\left(2k\pi/n\right)}$$

$$= \frac{e^2}{2a^2} \sum_{k=1}^{n-1} \frac{\sin\left(k\pi/n\right)}{1 - \cos\left(2\pi k/n\right)} = \frac{e^2}{2a^2} \sum_{k=1}^{n-1} \frac{\sin\left(k\pi/n\right)}{2\sin^2\left(k\pi/n\right)}$$

$$= \frac{e^2}{4a^2} \sum_{k=1}^{n-1} \csc\left(k\pi/n\right) \equiv \frac{e^2}{4a^2} S_n \,. \qquad \textbf{(1)}$$

He then considered the restoring forces on the rotating charges when slightly perturbed, and found the necessary conditions for them to oscillate with frequencies $q$ about their equilibrium positions; for the specific case of small displacements in the plane of rotation, these frequencies were given by the roots to a set of $n$ equations:[19]

$$\left(\frac{3e^2}{4a^3} S_k + L_k - L_0 - mq^2\right)\left(N_0 - N_k - mq^2\right) = \left(M_k - 2m\omega q\right)^2 \qquad k = 0,1,2\ldots(n-1) \quad \textbf{(2)}$$

Through a series of lengthy calculations, Thomson showed one-by-one that for $n < 6$, the $2n$ solutions to these equations would all be real numbers (corresponding to stable oscillations), as long as the rotational speed of the ring exceeded some minimum value. However, in the case of $n = 6$, the $k = 3$ equation from (2) reads as:

---

[18] Thomson did not explicitly derive this formula in his paper, but the derivation is presented here as an example of the numerous trigonometric quantities that appeared throughout.

[19] The indexed quantities $L$, $M$ & $N$ appearing in (2) represent various sums of trigonometric functions, similar to the quantity $S$ defined in (1).

$$\left(-\frac{14 - 8\sqrt{3}}{8}\frac{e^2}{a^3} - mq^2\right)\left(58\frac{e^2}{8a^3} - mq^2\right) = 4m^2\omega^2q^2 \,. \tag{3}$$

For this equation, one of the roots for $q^2$ will be negative (making $q$ imaginary), resulting in a pair of solutions with exponential time dependence. Thus, a ring of six charges would be mechanically unstable, no matter its angular velocity, *unless* an additional negative charge were placed at the center of the sphere.[20] With six charges in a ring and a seventh located at the center, the roots to the suitably modified frequency equations would all be positive in $q^2$, and therefore oscillatory. He further found that if the number of outer charges increased beyond eight, more and more interior charges would be required for the dynamic stability of the system. [Table I]

| $n$ | 5 | 6 | 7 | 8 | 9 | 10 | 15 | 20 | 30 | 40 |
|---|---|---|---|---|---|---|---|---|---|---|
| $p$ | 0 | 1 | 1 | 1 | 2 | 3 | 15 | 39 | 101 | 232 |

**Table I.** Number of charges $n$ contained in a single atomic ring, and the minimum number of interior charges $p$ required for the dynamic stability of that ring. [From Thomson 1904, p. 254.] The last entries represent the largest number of electrons in a single atom mentioned by Thomson in this paper.

Thomson thought these interior charges would (like Mayer's magnets) also arrange themselves into closed, stable orbits, and that atoms with $n > 8$ ought to be made up by a series of concentric rings. The number of charges would vary from ring to ring, with the greatest number in the outermost ring, and a decreasing number of charges occupying each of the inner rings. [Table II] Thomson then proceeded to argue how the periodicity of these configurations (and their resultant properties, determined by the number occupation and stability of the rings) closely matched the periodicity of the natural elements. Noticing, for example, that the interior configuration for an atom with $N = 60$ charges would be identical to that of

---

[20] A single negative charge added at the center introduced a *positive* quantity to the first term in parentheses of (2) & (3), so that the offending term in (3) would be manifestly positive. [Thomson 1904, p. 251-2]

*N* = 40, Thomson proposed that groups of such atoms [each one derived from the previous by the addition of a single ring, as with the *N* = 3, 11, 24, 40 & 60 atoms in Table II] would have similar chemical and spectral properties, and should therefore fall into the same vertical column in the Periodic Table. [Thomson 1904, p. 259]

| *N* | 3 | 11 | 15 | 20 | 24 | 30 | 35 | 40 | 45 | 50 | 55 | 60 |
|---|---|---|---|---|---|---|---|---|---|---|---|---|
| | 3 | 3 | 5 | 1 | 3 | 5 | 1 | 3 | 4 | 1 | 1 | 3 |
| | | 8 | 10 | 7 | 8 | 10 | 6 | 8 | 10 | 5 | 7 | 8 |
| | | | | 12 | 13 | 15 | 12 | 13 | 14 | 11 | 12 | 13 |
| | | | | | | | 16 | 16 | 17 | 15 | 16 | 16 |
| | | | | | | | | | | 18 | 19 | 20 |

**Table II.** Sampling of charge arrangements for a number of different atoms, from the inner rings (upper rows) to the outer rings (lower rows), where *N* represents the total number of charges in each atom. The *N* = 3, 11, 24, 40 & 60 atoms would form a single vertical column in the Periodic Table, each member of the group derived from the previous member by the addition of a single ring. [Adapted from Thomson 1904, p. 257.]

In a similar vein, atoms in the group *N* = 59 – 67 each had 20 charges in their outermost rings. [Table III] The greater the number of interior charges, the more stable the outer rings would be, making the outer ring for *N* = 59 the least stable of the group. Thomson characterized this ring as "very near the edge of stability," so that it might easily *lose* one of its charges, perhaps due to some external force. Having done so, it would then behave like a strongly electropositive atom and immediately attract any negative charges in its vicinity, so that it would be unable to maintain its charged state, and act in the long term as though it were chemically inert. [Thomson 1904, p. 261]

| $N$ | 59 | 60 | 61 | 62 | 63 | 64 | 65 | 66 | 67 |
|---|---|---|---|---|---|---|---|---|---|
| | 2 | 3 | 3 | 3 | 3 | 4 | 4 | 5 | 5 |
| | 8 | 8 | 9 | 9 | 10 | 10 | 10 | 10 | 10 |
| | 13 | 13 | 13 | 13 | 13 | 13 | 14 | 14 | 15 |
| | 16 | 16 | 16 | 17 | 17 | 17 | 17 | 17 | 17 |
| | 20 | 20 | 20 | 20 | 20 | 20 | 20 | 20 | 20 |

**Table III.** Distribution of charges from inner rings (top rows) to outer rings (bottom rows) for the group of atoms containing $N = 59 - 67$ charges. Each has 20 charges in its outermost ring, placing them all in a single horizontal row of the Periodic Table. [From Thomson 1904, p. 258]

With additional interior charges, successive atoms in this group would be less and less electropositive,[21] eventually to the point where they could *accept* one or two extra charges while still maintaining stability; these atoms would have an increasingly electronegative character. Maximum stability would occur for $N = 67$, which wouldn't be able to retain any extra charges for long because $N = 68$ corresponded to a quasi-stable outer ring of 21 charges. Similar to $N = 59$, this atom would quickly lose any additional charge it had gained, and also act as though it had no valency. Thomson argued that the atomic properties predicted by his model were reflected in a series of known elements, when arranged as:

| He | Li | Be | B | C | N | O | F | Ne |
|---|---|---|---|---|---|---|---|---|
| Ne | Na | Mg | Al | Si | P | S | Cl | Arg |

The first and last elements in each row have no valency, the second are monovalent electropositive, the second-to-last monovalent electronegative, and so on. [Thomson 1904, p. 260-62]

He also drew attention to there being *two* sets of vibrations associated with his model: one corresponding to the angular rotation of the rings, the other arising from the oscillation of the charges about their equilibrium positions. Though he'd taken great pains to derive formulas for each of these allowed frequencies (for $n \leq 6$), he made no mention of any attempt at quantitative comparisons with atomic spectra. After all, his model placed no constraints on the angular velocities of the rings (other than exceeding some minimum value), the charge and mass of an electron were not separately known with precision, and he had no way of determining the absolute number of charges contained within an atom.[22]

---

[21] The outer rings would become *more* stable, so that they would be *less* likely to lose a charge and behave like an electropositive atom.

Thomson waited until the very last of his paper to finally address the problem of radiative instability. As mentioned previously, classical electrodynamics predicts that any accelerated charge will radiate energy in the form of electromagnetic waves. The kinetic energy of a system of point charges in periodic motion would therefore be dissipated by the resulting radiation until exhausted. Thomson imagined a ring of charges initially rotating with sufficient angular speed, but that

> "... in consequence of the radiation from the moving corpuscles, their velocities will slowly – very slowly – diminish; when, after a long interval, the velocity reaches the critical velocity, there will be what is equivalent to an explosion of the corpuscles [...] The kinetic energy gained in this way might be sufficient to carry the system out of the atom, and we should have, as in the case of radium, a part of the atom shot off. In consequence of the very slow dissipation of energy by radiation the life of the atom would be very long." [Thomson 1904, p. 265]

Radiative instability might therefore be a mechanism for the production of beta radiation, but Thomson provided no further rationale here for why this would be a "slow" process. Justification *can* be found, however, in his paper from the previous year, detailing the fields produced by symmetrically arranged point charges constrained to rotate in a circle. His conclusion had been that if the velocities of the particles were small compared to the speed of light, the rate of radiative energy loss would dramatically *decrease* as the number of discrete charges on the ring *increased*. [Thomson 1903, p. 681; see Table IV] This result may at first seem counterintuitive, but recall that a steady loop of current produces time-*independent* fields, and does not radiate electromagnetic waves. Thomson found that the circular motion of *discrete* charges would produce oscillatory fields that, due to the spatial symmetry, had an increasing tendency to cancel out as the number of orbiting charges increased.

Thomson mentioned atomic mass only once in his 1904 article, in connection with the number of charges in an atom: "We suppose that the mass of an atom is the sum of the masses of the corpuscles it contains, so that the atomic weight of an element is measured by the number of corpuscles in its atom." [Thomson 1904, p. 258] This implied that the positive charge contributed nothing to the atomic mass. With no experimental method for determining whether this was indeed true, the question remained open of whether or not an atom actually contained many thousands of electrons – perhaps all but one or two of them were too tightly bound to be stripped in a standard discharge lamp.

| Number of Particles | Radiation from each particle. | |
|:---:|:---:|:---:|
| | $a\omega = c/10$ | $a\omega = c/100$ |
| 1 | 1 | 1 |
| 2 | $9.6 \times 10^{-2}$ | $9.6 \times 10^{-4}$ |
| 3 | $4.6 \times 10^{-3}$ | $4.6 \times 10^{-7}$ |
| 4 | $1.7 \times 10^{-4}$ | $1.7 \times 10^{-10}$ |
| 5 | $5.6 \times 10^{-5}$ | $5.6 \times 10^{-13}$ |
| 6 | $1.6 \times 10^{-7}$ | $1.6 \times 10^{-17}$ |

**Table IV.** Average radiation per particle for $n \leq 6$ charges rotating at constant angular speed $\omega$, at a distance $a$ from the center, relative to the case of a single orbiting charge (taken as unity). The middle column represents charges moving at 1/10 the speed of light, the other at 1/100 the speed of light. [From Thomson 1903, p. 681.]

## 2.3 The number of corpuscles in the atom

In a paper published just two years later, Thomson used a modified, static version of his model to propose several methods for determining the total number of negative charges in an atom, based on experimental data for three different phenomena: the dispersion of light and the scattering of X-rays by dilute gases; and the absorption of β-rays when traversing matter. In each case he was led to the same conclusion: the number of charges in an atom should be on the same order of magnitude as its atomic weight.

> "It will be seen that the methods are very different and deal with widely separated physical phenomena; and although no one of the methods can, I think, be regarded as quite conclusive by itself, the evidence becomes very strong when we find that such different methods lead to practically identical results." [Thomson 1906, p. 769]

Thomson first derived a formula for the refractive index of a dilute monatomic gas, by considering the effect of a sinusoidal electric field on a uniformly charged sphere containing $n$ negative point charges of mass $m$ (distributed uniformly throughout the atom). The external field would cause the positive and negative charges to be displaced in opposite directions, thereby polarizing the atoms and increasing the relative permittivity of the gas, as measured by the index of refraction $\mu$. He immediately conceded that the positive charge must contribute some amount $M$ to the total mass of the atom, otherwise the induced polarization would have no frequency dependence, making dispersion impossible.[23] His final expression for the index of refraction was written in terms of the quantities $P_0$ (the net volume occupied by atoms per cubic centimeter of gas), $N$ (the number of atoms per unit volume) and the incident wavelength $\lambda$:

---

[23] Thomson considered only wavelengths that were long compared to atomic distances, so that the external field could be taken as uniform over the length of an atom. If the inertial mass of either the positive or negative charges were zero, there would be no frequency dependence in the response of the atom to the time-varying electric field, and the maximum polarization would depend only on the amplitude of the incident wave.

$$\frac{\mu^2-1}{\mu^2+2} = P_0 + P_0^2 \frac{M}{E'} \frac{m}{e'} \frac{1}{N(M+nm)} \frac{3\pi}{\lambda^2},$$ **(4)**

where $E'$ and $e'$ are respectively the total charge of the positive sphere and a negative corpuscle, expressed in electromagnetic units. [Thomson 1906, p. 771]

Thomson was unaware of any experimental data on the dispersive power of monatomic gases, but he used a finding by Lord Rayleigh (that the dispersion of helium is of the *same order* as a diatomic gas) to justify comparing his theory with data published by E. Ketteler on the refractive index of molecular hydrogen. At atmospheric pressure, Ketteler's measurements yielded the expression:

$$\frac{\mu^2-1}{\mu^2+2} = \frac{1}{3}\left\{2.8014 \times 10^{-4} + \frac{2 \times 10^{-14}}{\lambda^2}\right\}.$$ **(5)**

Comparing this with (4), and using $e'/m = 1.7 \times 10^7$ and $Ne' = 0.8$, Thomson found that, approximately:

$$\frac{M}{M+nm} \frac{1}{n} \simeq 1.$$ **(6)**

Therefore, the number of negative charges $n$ per hydrogen atom should not be much different from one, and the mass of the positive charge $M$ must be large compared to $nm$, the total mass of all the negative charges. [Thomson 1906, p. 771]

The second experimental method, concerning the scattering of X-rays, was given considerably less attention by Thomson in this paper, because the theory had already been developed in his recent book, "Conduction of Electricity through Gases."[24] He compared a previously derived expression for the relative amount of X-rays scattered by a gas of $N$ electrons per cubic centimeter with data from C. G. Barkla, who had measured the ratio of scattered to incident energy for X-rays passing through air to be $2.4 \times 10^{-4}$. Setting these equal:

$$\frac{8\pi}{3} \frac{Ne^4}{m^2} = 2.4 \times 10^{-4},$$ **(7)**

---

[24] Published by Cambridge University Press in 1903, second edition appearing in 1906.

and putting in $e/m$ = 1.7×10$^7$ and $e$ = 1.1×10$^{-20}$, he found that the number of negative charges in each molecule of air should be around 25, very close to the molecular mass of nitrogen at approximately 28. [Thomson 1906, p. 772-3]

The third method involved the absorption of β-rays, and is of particular interest as an initial theoretical treatment by Thomson of a topic that would play an important role in deciding between his atomic model and Rutherford's nuclear atom – the scattering of high-velocity electrons by matter.  Thomson believed the loss of kinetic energy for β-particles in an absorbing substance, and their deflection when passing through thin sheets of metal, could *both* be explained as resulting from a series of multiple encounters with the individual atoms in the material. [Thomson 1906, p. 773] In order to greatly simplify the problem, he assumed that the change in energy for a β-particle deflected by a *single* atom would be small (corresponding to a small deflection), and due *only* to interactions with the negative point charges, again modeled as being at rest and uniformly distributed.  For a succession of such encounters, his calculations indicated that the number of transmitted particles would decrease exponentially[25] with the thickness of the scatterer, as characterized by the coefficient of absorption $\lambda$.   Solving for $\lambda/\delta$ (the ratio of the absorption coefficient to the density of the medium, which Thomson knew to be approximately constant for any absorbing material), he arrived at:

$$\frac{\lambda}{\delta} = 4\pi \frac{e^4}{m^2} \frac{n}{M} \frac{V_0^4}{V^4} \log\left( \frac{1}{2} \frac{aV^2}{V_0^2} \frac{m}{e^2} - 1 \right). \qquad \textbf{(7)}$$

Here, $V$ is the velocity of the incident particles, $V_0$ the speed of light, and the undetermined quantity $a$ is a length comparable to the distance between atomic charges.  Since the logarithmic term would be slowly varying, he reasoned that $n/M$ (the ratio of the number of charges to the total mass of the atom) should also be roughly constant. [Thomson 1906, p. 773]

---

[25] There was considerable debate at the time, and for long after, whether the absorption law for monoenergetic electrons should be linear or exponential.  For details, see Chapter 1 of Franklin, 2001.

To find an expression for the actual number of corpuscles per atom, he set $V = 1.7 \times 10^{10}$ (for β-particles from a uranium source, according to Becquerel) and $\lambda/\delta = 7$ (for copper and silver, according to Rutherford), and substituted $e/M' = 10^4$ (the charge-to-mass ratio for ionized hydrogen):

$$n = \frac{M}{M'} \cdot \frac{1.4}{\log\left(\dfrac{mV^2}{V_0^2}\dfrac{a}{e^2} - 1\right)}. \qquad \textbf{(8)}$$

Once again, the number of charges per atom was roughly proportional to the atomic weight.  One final interpretation of this result was that $e/M$ for the positive charge must be of the order $10^4$, over 1000 times smaller than the charge-to-mass ratio for an electron. [Thomson 1906, p. 774]

Thomson stated only one "obvious" objection to the number of charges in an atom being as low as its atomic weight: the large number of spectral lines exhibited by excited atoms in a magnetic field (the Zeeman effect).  Thomson admitted this objection might be well founded if the charges excited by a flame, or by an electrical discharge, were simply vibrating within a normal atom, but he argued there was no evidence of this actually being the case, for the free electrons in an ionized gas might be capable of making a number of *temporary* combinations, which could also explain the existence of so many spectral lines. [Thomson 1906, p. 774]

It seems that Thomson would only run into difficulties if he tried to reconcile his model with the widely accepted view of every spectral line corresponding to an oscillating atomic electron, but he was beginning to find self-consistent results when comparing his theoretical predictions with data from scattering and absorption experiments.  By 1910, Thomson had generalized his earlier calculations to take into account the deflection of β-particles by encounters with *both* the positive and negative charge inside an atom.  Citing a simple probability argument by Lord Rayleigh,[26] he set the mean total deflection $\phi_m$ of a high-energy β-particle, resulting from $n$ atomic encounters (each producing a small average deflection $\theta$) equal to

---

[26] He cited Rayleigh, *Theory of Sound*, Second Edition, Vol. I, p. 35 (1894).  This is a "random walk" argument for the average resultant of $n$ displacements of arbitrary phase and constant amplitude $\theta$.

$\sqrt{n} \cdot \theta$, which translates into $\phi_m = \sqrt{N\pi R^2 t} \cdot \theta$ for scattering in a plate of thickness $t$ having $N$ atoms of radius $R$ per unit volume. The remaining task would be to find a theoretical value for $\theta$ using the assumptions of his latest scattering models. [Thomson 1910, p. 465]

He first derived an expression for the mean deflection $\theta_1$ due to intra-atomic encounters with the $N_0$ negatively charged point particles per atom (again, distributed uniformly). This led to the expression

$$\theta_1 = \frac{16}{5} \frac{e^2}{mv^2} \frac{1}{R} \sqrt{\frac{3N_0}{2}} \tag{9}$$

for β-particles with an initial kinetic energy of $1/2mv^2$. As for the positive charge (equal to $eN_0$ for a neutral atom), if it took the form of a uniform sphere with radius $R$, the mean deflection from a single encounter would be

$$\phi_1 = \frac{\pi}{4} \frac{e^2}{mv^2} \frac{N_0}{R} . \tag{10}$$

However, if the positive charge in an atom happened to *instead* be made up from tiny particles (distributed in the same way as the negative charges), the mean deflection would become

$$\phi_2 = \theta_1 \cdot \sqrt{1 - \left(1 - \frac{\pi}{8}\right)\sigma^{1/3}} , \tag{11}$$

where $\sigma$ is the ratio of the net volume occupied by positive charges to the volume of an atom. The total mean deflection per atomic encounter would then be either $\phi_m = \left(\theta_1^2 + \phi_1^2\right)^{1/2}$ or $\phi_m = \left(\theta_1^2 + \phi_2^2\right)^{1/2}$, depending on how the positive charge was distributed. [Thomson 1910, p. 466-7]

There are two points worth noticing about these expressions. First, if $\sigma$ turned out to be negligible, then $\phi_2$ would be exactly equal to $\theta_1$ (the mean deflection due to negative point charges, as would be expected if the positive and negative charges existed in similar states). Because $\sigma$ has a maximum value of one, the two quantities would at any rate be of the same order of magnitude. Second, the average deflection $\phi_1$ due to a *uniform sphere* of positive charge would be directly proportional to $N_0$ (the number of charges in a neutral atom), whereas $\phi_2$ and $\theta_1$ are both proportional to $\sqrt{N_0}$ . This implies – for atoms with a large value of $N_0$ – that the dominant contribution to the total scattering would be from the positive sphere, and not the negative point charges. This fact will turn out to be relevant for comparisons made by Rutherford in 1911 on the scattering predicted for single encounters with a dense atomic core.

Whether the positive charge existed in the form of point particles or as a uniform sphere, an important prediction of both possibilities was that $\phi_m / t^{1/2}$ (the ratio of the mean deflection angle to the square root of the thickness of the scatterer) should be inversely proportional to the kinetic energy of the incident particles, and equal to a constant if the beam were energetically homogeneous. This was confirmed by J. A. Crowther (also in 1910) using very thin sheets of metal, and a beam of β-particles with velocities that differed by less than one percent.[27] As part of his experiments, Crowther determined for several materials the thickness $t_m$ that would reduce by half the number of particles scattered inside a fixed angle $\phi$, which he then used to calculate the total number of charges per atom $N_0$ according to both of Thomson's proposals. [Table V]

Assuming the positive charge to be uniformly distributed in a sphere of atomic radius (Theory A) resulted in a nearly constant value for all the scattering materials of about three electrons per unit of atomic weight. The other assumption, that the positive charge existed as point particles distributed uniformly throughout a spherical volume (Theory B), gave a ratio of $N_0$ to atomic weight that increased dramatically for heavier scatterers, contradicting Thomson's earlier conclusion that this ratio should be approximately constant, and on the order of one. Based on these results, Crowther decided that

---

[27] The uniformity of the beam was essential for confirming this aspect of Thomson's theory. Crowther went into explicit detail regarding his methods for achieving a homogeneous beam, saying that certain subtleties had been overlooked by past experimenters. He first passed the particles through a magnetic field, then selected for a small range of velocities with a tiny aperture. Crowther pointed out that an overly sized aperture would admit a greater range than what would naively be expected. [Crowther, p.229-33]

"...the positive electricity in an atom is not in a state comparable to that of the electron, but ... occupies such comparatively large volumes as to be capable of being considered as uniformly distributed over the whole atom." [Crowther 1910, p. 240]

| Element. | Atomic weight. | $\phi/\sqrt{t_m}$ | $N_0$ | | $N_0/$Atomic weight | |
|---|---|---|---|---|---|---|
| | | | A. | B. | A. | B. |
| Carbon* | 12 | 2.0 | 40 | 44 | 3.32 | 3.7 |
| Aluminum | 27 | 4.25 | 83 | 156 | 3.07 | 5.8 |
| Copper | 63.2 | 10.0 | 181 | 765 | 2.87 | 12.0 |
| Silver | 108 | 15.4 | 320 | 2080 | 2.96 | 19.2 |
| Platinum | 194 | 29.0 | 605 | 6500 | 3.12 | 33.5 |

**Table V.** Ratio of atomic charges $N_0$ to atomic weight for several scattering materials, calculated on the assumption of either a uniform sphere of positive charge (A), or an "electronic" distribution of positively charged point particles (B). The measurements for carbon involved a substance ("caoutchouc") that was 90% carbon and 10% hydrogen. [From Crowther 1910, p. 239]

Thomson's model and Crowther's experimental data were together producing numbers that were consistent with scattering from a spherical charge distribution, though they were off (in retrospect) by around a factor of three. But recall that for large values of $N_0$ (true for most of the materials tested by Crowther), a theory based on a uniformly distributed sphere of charge would predict the scattering of high velocity β-particles to be dominated by interactions with the positive charge, as expressed in equation (10). When estimated by Rutherford in 1911, he found that the average scattering angle would be dominated by *single* encounters with a massive nuclear charge, and be exactly *three times greater* than for a series of small deflections due to *multiple* encounters with a Thomson atom. [Rutherford 1911, p. 678; see below] Given the limitations of the currently available experimental methods, the much more massive α-particles being studied by Rutherford and Hans Geiger were better suited than β-particles for probing the internal structure of atoms through scattering experiments, in part because their relatively high kinetic energies meant that interactions with the atomic electrons could be safely neglected.

# 3 The nuclear atom of Ernest Rutherford

## 3.1 Fundamental properties of α-particles

When Rutherford gave his first quantitative description[28] of the scattering of α-particles by matter, he was quick to point out how this phenomenon had important implications for the nature of atoms:

> "…it can easily be calculated that the change of direction of $2^0$ … would require over that distance an average transverse electric field of about 100 million volts per cm. Such a result brings out clearly the fact that the atoms of matter must be the seat of very intense electrical forces – a deduction in harmony with the electronic theory of matter." [Rutherford 1906a, p. 145]

Among the primary objectives of the experiments reported by Rutherford in 1906 was to determine the relative amount of energy lost by α-particles when passing through absorbing materials of various thicknesses (in this case, thin layers of aluminum sheets). This could only be accomplished using an energetically homogeneous α-source, and Rutherford had recently developed a special technique for creating one.[29]

His experiments consisted of placing the apparatus inside a uniform magnetic field that was reversed in direction every 10 minutes over a period of two hours; after passing through a narrow slit in a moveable screen, the α-particles were detected with photographic plates. [Fig. 3] Half of the α-source was covered in thin layers of aluminum, and the other half left essentially uncovered, so that Rutherford

---

[28] Rutherford remarked in a paper published in January 1906 (dated 15 November 1905) that he'd observed a definite scattering of α-particles when passing through air (but no quantitative measurements), and that experiments were currently underway to see whether this scattering also occurred for α-particles traversing solids. [Rutherford 1906b, p. 174]

[29] In 1905, Rutherford had criticized several experiments (by himself and others) for using thick layers of a mixture of radium and its decay products as an α-source. Rutherford could create a sufficiently homogeneous source by exposing a negatively charged wire for several hours to *radium emanation* (radon gas). The thinness of the active layer deposited on the wire ensured the particles would all escape from the source with the same velocity. With the various radium products having different half-lives, a dominant radiation source could be selected according to the time elapsed after exposing the wire – in the case of radium C (bismuth-214), this amounted to waiting at least 15 minutes after exposure before beginning an experiment. [Rutherford 1905a, p. 165]

could directly compare how the two types of particles were deflected by the velocity-dependent force of the applied magnetic field. The time-alternating field caused two sets of bands to be produced, and the distances measured between the edges of the bands in each set were inversely proportional to the velocity of the particles having passed through the absorbing layers, relative to the ones that had not. [Rutherford 1906a, p. 136]

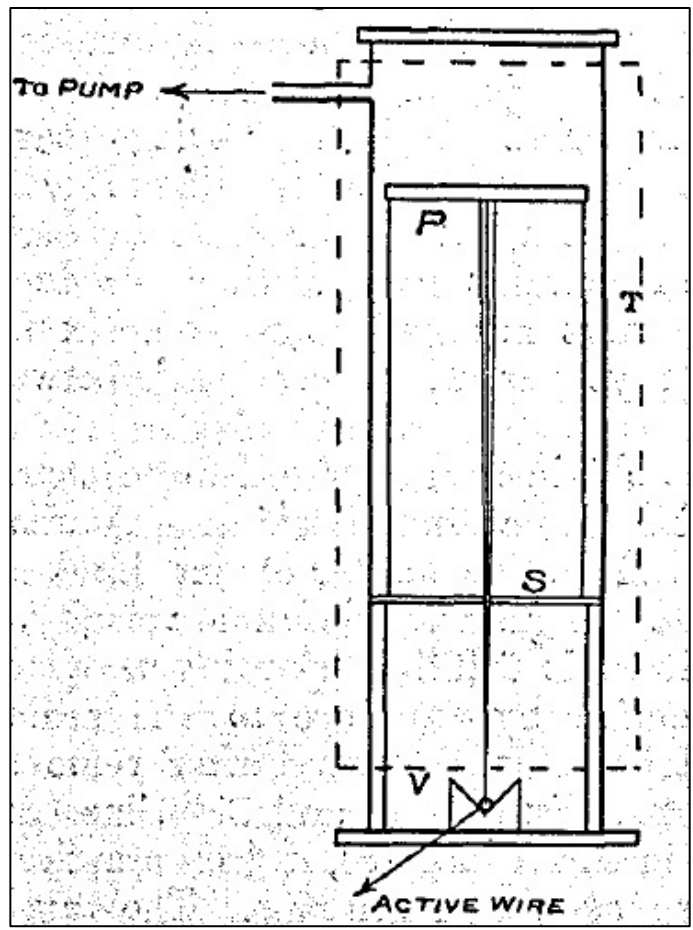

**Figure 3.** Apparatus used by Rutherford to study the magnetic deflection of α-particles after traversing thin sheets of aluminum. In an evacuated brass tube *T*, an active source was placed in the slot *V* at 2 cm from a narrow slit *S* in an adjustable screen. After passing through the slit, the particles were detected by the photographic plate *P*. The external magnetic field was non-zero and roughly uniform in the region contained by the dashed lines. [From Rutherford 1905a, p. 166]

Rutherford had initially used a microscope to directly measure the traces left by the α-particles on the photographic plates, but soon realized this was only effective when the traces were pronounced, and that the method "proved very trying to the eyes." He was able to make more accurate measurements by projecting the images onto a large cardboard screen, then marking the edges of the visible bands with a sharp pencil. [Rutherford 1906a, p. 136-7]  In what was essentially an aside to the main results of his paper, he commented on how the deflections he'd observed for α-rays were relatively small compared to the scattering of β-rays, which should be expected considering the enormous differences in their momenta and kinetic energies.  Still, the degree of scattering by the absorbing material was significant enough to be noteworthy, in particular because the least energetic α-particles were the most likely to escape detection.

> "From measurements of the width of the band due to the scattered α rays, [some of them] have been deflected from their course by an angle of about $2^0$. It is possible that some were deflected through a considerably greater angle; but, if so, their photographic action was too weak to detect on the plate." [Rutherford 1906a, p. 144-5]

His experiments had indicated that an α-particle's ability to produce a photographic effect decreased more rapidly than its loss of kinetic energy, as there appeared to be a "critical velocity" below which it could no longer be detected by a photosensitive plate.[30]  He had similarly observed a reduction in the intensity of scintillations produced by a phosphorescent zinc sulfide screen after the α-particles had suffered some absorption.  Without a precise understanding of the mechanisms involved, Rutherford thought this result made sense if either method somehow required a minimum ionization power for the particles to effect a change in the detecting medium.[31] [Rutherford 1906a, p. 145-6]

According to Ernest Rutherford, the greatest question in radioactivity at the time was whether an α-particle could be identified as an atom of helium.  Working under the assumption that they carried a single unit of charge, he had previously

---

[30] Rutherford had found that 12 layers of aluminum foil resulted in a velocity reduction of 62%, but the α-particles could not be detected at all by the photographic method with 13 layers in place. [Rutherford 1906b, p. 170]

[31] Rutherford contemplated two explanations: the α-particle somehow lost its ionizing power when its velocity fell below ~40% of the incident beam (he could think of no obvious reason for why this should be so); or there was a sudden, rapid decrease in velocity in an absorbing medium when reaching this critical speed.  The scattering of the α-particles made definite conclusions impossible, and he anticipated further experiments to investigate the issue more closely. [Rutherford 1906a, p. 145-6]

estimated in 1905 the number of α-particles emitted per second from a radium sample, by measuring the ionization current they produced when passing through a magnetic field between two metal plates fixed at opposite potentials.[32]  After later obtaining a value for the α-particle's mass in separate experiments, Rutherford recognized that a charge of $2e$ would be required in order to match the charge-to-mass ratio for doubly-ionized helium.  This would mean having to cut his earlier estimate on the activity of radium by half. [Rutherford and Geiger 1908a, p. 141]

As such, it would be a tremendous accomplishment to measure this activity independent of any assumptions about the charge, for if the total number of particles emitted per second could be counted directly, the magnitude of their charge would be easily deduced.  By 1908, Rutherford and Hans Geiger had together managed to conclusively detect individual α-particles by magnifying the effect of their passage.  They set the potential between two metal plates *nearly* high enough to spark in air – the intense field accelerated any electrons liberated by a passing α-particle, and they in turn collided with other gas molecules to create an ionization cascade.[33]  The immediate change in potential across the plates was unmistakable.

"Any sudden rise of potential of the electrometer, for example that due to the entrance of an α-particle in the detecting vessel, then manifested itself as a sudden *ballistic* throw of the electrometer needle.  The charge rapidly leaked away and in a few seconds the needle was again at rest in its old position." [Rutherford and Geiger 1908a, p. 144; emphasis in the original]

Once this method had been shown to produce consistent numbers,[34] the two of them repeated the ionization current measurements reported by Rutherford in 1905, and this time found a charge of somewhere between $2e$ and $3e$.  If the charge

---

[32] Assuming unit charge, he estimated the activity to be $6.6×10^{10}$ particles per second from 1 gram of radium. [Rutherford 1905b, p.199] Compare half this value (corresponding to a charge of $2e$) with the definition of the curie ($3.7×10^{10}$ decays per second), which is based on the activity of 1 gram of radium-226.

[33] Note the similarities between this method and the design of the Geiger counter, invented in 1908 for the detection of ionizing radiation.

[34] The average value was $3.4×10^{10}$ particles per second for one gram of radium in equilibrium with its three α-producing decay products. [Rutherford and Geiger 1908a, p. 156]

had to be an integral multiple of the fundamental unit, the evidence was strongly in favor of the former over the latter, in particular because a charge of 2$e$ would correspond to the α-particle having an atomic weight of 3.84, very close to the atomic weight of helium at 3.96.

> "Taking into account probable experimental errors … we may conclude that *an α-particle is a helium atom*, or, to be more precise, *the α-particle, after it has lost its positive charge, is a helium atom.*" [Rutherford and Geiger 1908b, p. 172; emphasis in the original]

Having established a reliable method for counting individual α-particles, Rutherford and Geiger could now compare the number of scintillations they observed on a zinc sulfide screen with the number of particles they expected to be produced under similar conditions (the "calculated" values in Table VI). The agreement between the two sets of data gave them confidence that, within experimental error, the ionization and the scintillation methods would both generate the same numbers. [Rutherford and Geiger 1908a, p. 158]

| I. Calculated number of α-particles per minute. | II. Observed number of scintillations per minute. | III. Ratio of observed to calculated number. |
|---|---|---|
| 39 | 31 | 0.80 |
| 38 | 49 | 1.29 |
| 34 | 29 | 0,85 |
| 32 | 31 | 0.97 |
| 31 | 32 | 1.03 |
| 28 | 27 | 0.96 |
| 27 | 28 | 1.04 |
| 25 | 21 | 0.84 |
| 23 | 25 | 1.09 |
| 21 | 21 | 1.00 |
| Total number = 150 | | Average = 0.96 |

**Table VI.** Comparison of the expected number of α-particles produced per minute (I) with the number of scintillations actually observed on a zinc sulfide screen (II). The ratio of the two numbers (III) showed they were roughly equivalent. [From Rutherford and Geiger 1908a, p. 158]

## 3.2    The angular dependence of α-scattering

At Rutherford's suggestion, Hans Geiger set about investigating the angular dependence of α-scattering using the scintillation method of detection. The crux of a short paper authored by Geiger in 1908 was to unambiguously show that the scattering of α-particles by matter was a real phenomenon, and not an artifact of previous experimental methods.[35]  The relatively quick rate of decay for a radium C source (prepared by a method similar to the one established by Rutherford in 1905) made it difficult to obtain definite results, so Geiger instead used a highly active source of radon gas, held at low pressure inside a tapered glass tube that was sealed at the open end with a thin sheet of mica. [Geiger 1908, p. 175]

As they traversed a 2 meter long evacuated cylinder, the particles passed through a narrow slit before impinging on a small phosphorescent screen. [Fig. 4] In a dimly lit area, Geiger sat behind the screen and counted the number of scintillations produced using a microscope set at 50x magnification. He detected hardly any particles outside the geometrical image of the slit when the tube was fully evacuated, but saw definite scattering when he first placed one leaf of gold foil over the slit, and again when he added a second layer. [Fig. 5]

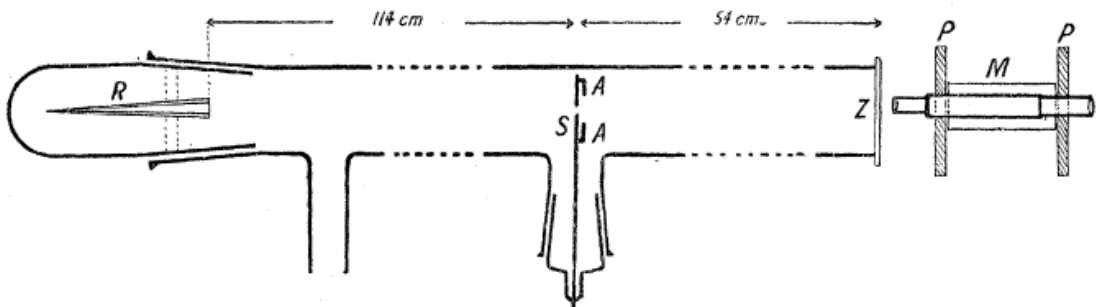

**Figure 4.** The first apparatus used by Geiger to study α-scattering.  In a 2 m evacuated glass tube, the α-particles from the source *R* passed through a narrow slit *S*, producing an image on the phosphorescent screen *Z*.  The slit was first left open, then covered with one, and then two metal foils. The microscope *M* could be adjusted to move across the screen. [From Geiger 1908, p. 174]

---

[35] There had been conflicting interpretations of the experimental results, in part because of the uncertainty introduced by the presence of a magnetic field; Geiger cited articles by Kucera and Masek (1906), W. H. Bragg (1906), L. Meitner (1907) and E Meyer (1907). [Geiger 1908, p. 174] In contrast to Rutherford's absorption experiments, Geiger did not use a magnetic field in his scattering experiments.

**Figure 5.** Number of particles detected per minute (in vacuum) versus distance from the center of the detecting screen, with the slit between the source and detector uncovered (A); also with one layer (B) and two layers (C) of gold leaf placed over the slit. [From Geiger 1908, p. 176]

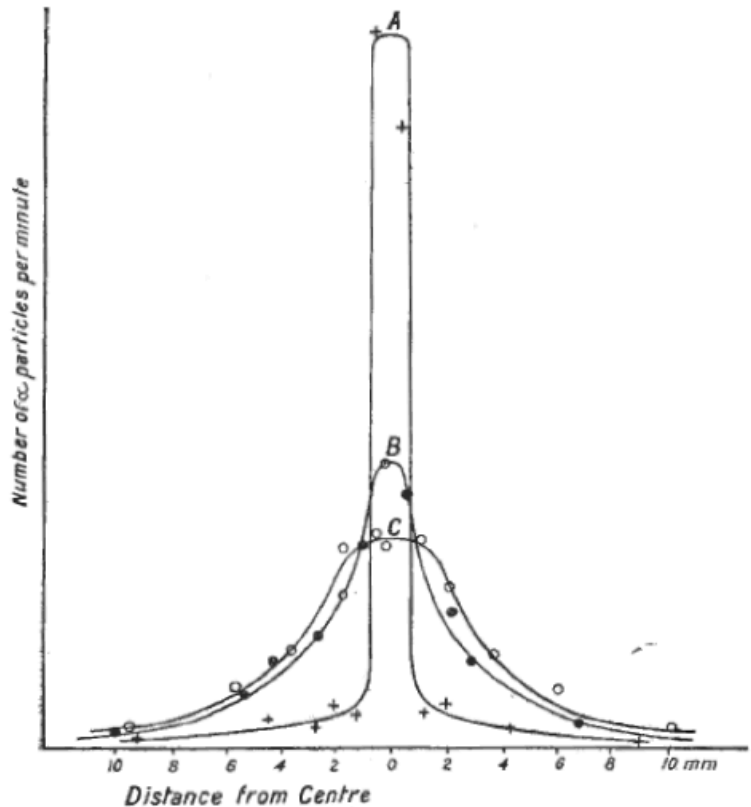

Other physicists of the time had observed that β-rays directed at a metal plate resulted in radiation being emitted from the same side of the plate as the incident particles, which was initially assumed to be a secondary effect. But recent experiments had shown that the observed radiation was actually comprised of primary particles that had been scattered to the extent of being reflected backwards. [Geiger 1909, p. 495] This was likely the motivation for Rutherford in early 1909 to suggest that Geiger, together with Ernest Marsden, attempt to see if α-particles would be reflected in a similar manner. In Marsden's own words:

> "One day Rutherford came into the room where [Geiger and I] were counting α-particles ... turned to me and said, 'See if you can get some effect of α-particles directly reflected from a metal surface.' [...] I do not think he expected any such result." [Quoted in Pais, p. 123]

Later that year, Geiger and Marsden reported having counted α-particles reflected off of several types of metals, and also the number reflected by different thicknesses of gold (which was readily available to them in the form of uniformly thin foils). In these experiments, a lead plate was placed between the α-source and the detecting screen, so that the only path available from source to detector was by reflection off a sheet of metal. [Fig. 6] They first showed that the amount of reflection increased according to the atomic weight of the reflector. [Geiger and Marsden, p. 495-7; see Table VII]

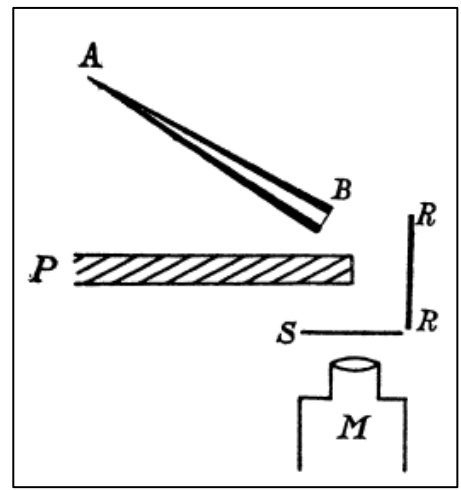

**Figure 6.** Setup used by Geiger and Marsden to study the reflection of α-particles from a metal surface. The lead plate *P* is situated between the α-source *AB* (an active glass tube) and the detecting screen *S* (observed with microscope *M*). The only path from source to detector was by reflection off the metal plate *RR*. [From Geiger and Marsden, p. 496]

| 1. Metal. | 2. Atomic weight, A. | 3. Number of scintillations per minute, Z. | 4. A/Z. |
|---|---|---|---|
| Lead | 207 | 62 | 30 |
| Gold | 197 | 67 | 34 |
| Platinum | 195 | 63 | 33 |
| Tin | 119 | 34 | 28 |
| Silver | 108 | 27 | 25 |
| Copper | 64 | 14.5 | 23 |
| Iron | 56 | 10.2 | 18.5 |
| Aluminium | 27 | 3.4 | 12.5 |

**Table VII.** Comparison of the number of α-particles reflected by various types of metals, which increases with the increasing atomic weight of the scatterer. The anomalous data for lead was attributed to suspected impurities in the metal. [From Geiger and Marsden, p. 497]

They next found that the number of scintillations they observed depended on the thickness of the reflector, in a way that was similar to reflection experiments using β-rays. [Fig. 7]   This was taken as clear evidence for extreme scattering occurring not only at the surface, but also within the volume of the material.  Their final measurements estimated the relative amount of large-angle scattering by a platinum reflector when the source was a known quantity of radium C; under these conditions, that number was around 1 in 8000. [Geiger and Marsden, p. 497-9]

"If the high velocity and mass of the α-particle be taken into account, it seems surprising that some ... can be turned within a layer of $1.65 \times 10^{-5}$ cm. of gold through an angle of $90^0$, and even more.  To produce a similar effect by a magnetic field, the enormous field of $10^9$ absolute units would be required."
[Geiger and Marsden, p. 498]

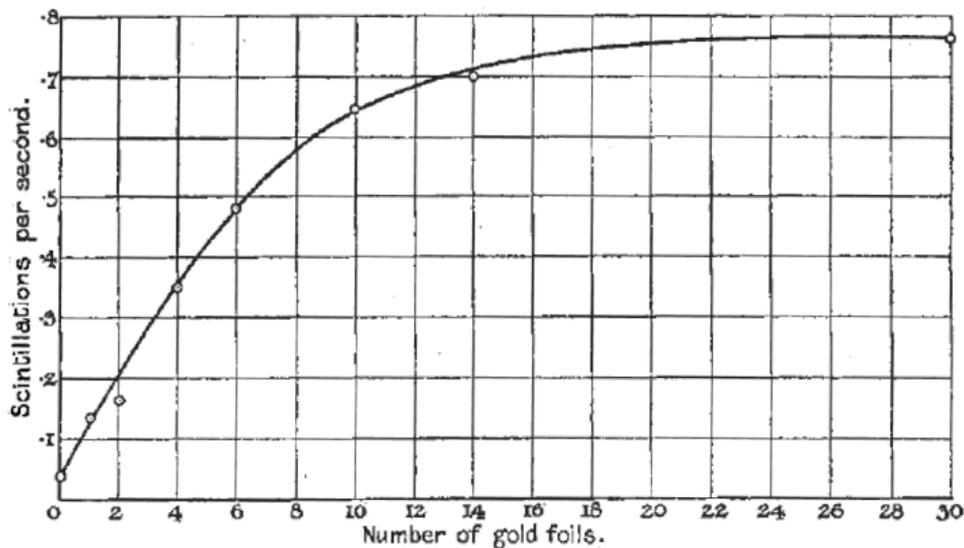

**Figure 7.** Number of scintillations per second, according to the thickness of the reflecting gold foil.  The first point of measurement (zero thickness) is reflection from a plate of glass; subsequent measurements placed increasing layers of gold on top of the glass. [Geiger and Marsden, p. 498]

By 1910, Geiger had amassed a substantial amount of scattering data by systematically varying the type and thickness of the scattering material, but for reliable measurements to be made he had required a better (closer to homogeneous) α-source than a low-pressure tube of radon gas sealed with mica. He instead placed an unsealed tube of glass inside a chamber filled with radon gas at near-atmospheric pressure for three hours, then evacuated the chamber to leave behind a thin layer of active deposit on the inner walls of the tube. [Geiger 1910, p. 492-3]

Once he'd determined that the peak scattering angle[36] increased linearly with the stopping power of the material, [Fig. 8] Geiger could argue that the most probable angle of deflection due to a single atomic encounter was directly proportional to the atomic weight $A$. To do this, he defined a *scattering coefficient K* as the most probable deflection angle for a thickness of material with stopping power equivalent to one centimeter of air. Geiger claimed that $K$ should be proportional to the square root of the atomic weight of the scatterer, because Bragg and Kleeman had shown in 1905 that the stopping power of a single atom was proportional to $A^{1/2}$. Geiger then defined a *relative atomic scattering* coefficient $K_0$ as the most likely deflection caused by a single atom (found by multiplying the scattering coefficient by the atomic weight, then setting this equal to unity for the case of gold). Taking the number of atoms encountered by an α-particle traversing materials with equivalent stopping power[37] to be *inversely* proportional to $A^{1/2}$, the ratios $K_0/A$ and $K/A^{1/2}$ were both expected to be roughly constant, in agreement with his experimental measurements. [Geiger 1910, p. 501-2; see Table VIII.]

---

[36] Geiger counted at semi-regular intervals over the course of about 70 minutes the total number of particles scattered into different areas on the screen, then corrected for the exponential decay in the activity of the source. The most probable angle corresponded to the peak in the curve found by plotting the number of scintillations at each angle. [Geiger 1910, p. 497]

[37] The total stopping power of a thin material was assumed to be the product of the stopping power per atom times the number of atoms per unit volume times the thickness. This product is the therefore the same for materials with equivalent stopping powers.

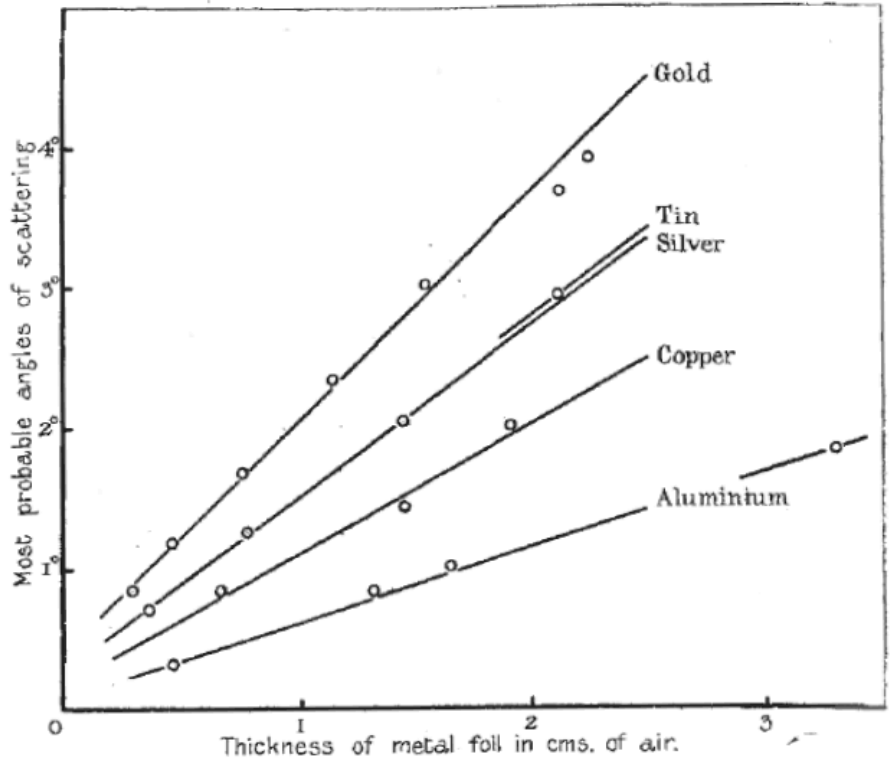

**Figure 8.** Most probable scattering angle for several materials, showing a linear dependence on the equivalent stopping power of the material (the thickness is expressed in terms of the material's equivalent stopping power in centimeters of air). [From Geiger 1910, p. 501]

| 1. | 2. | 3. | 4. | 5. | 6. |
|---|---|---|---|---|---|
| Scattering material. | Atomic weight, $A$. | Scattering coefficient, $K$ | $K/\sqrt{A}$ | Relative atomic scattering coefficient, $K_0$ | $\dfrac{K_0}{A} \times 10^2$ |
| Gold | 197 | 2.1 | 0.150 | 1.00 | 0.51 |
| Tin | 119 | 1.5 | 0.138 | 0.56 | 0.47 |
| Silver | 108 | 1.5 | 0.144 | 0.53 | 0.49 |
| Copper | 64 | 1.1 | 0.138 | 0.30 | 0.47 |
| Aluminium | 27 | 0.6 | 0.115 | 0.106 | 0.39 |

**Table VIII.** Most probable scattering angle $K$ (in degrees) for α-particles passing through a thickness of metal equivalent to the stopping power of 1 cm of air; and the amount of scattering $K_0$ for single atomic encounters, relative to the scattering produced by an atom of gold. The ratios $K/A^{1/2}$ and $K_0/A$ are roughly constant. [From Geiger 1910, p. 502]

Assuming the diameter of a gold atom to be about two angstroms,[38] Geiger estimated the average scattering angle caused by a single atomic encounter to be around 1/200th of a degree.  With this angle being so small, he concluded that the probability for an α-particle to be scattered into a very large angle from a series of multiple atomic encounters (*compound scattering*) would be vanishingly small, and certainly on a different order of magnitude than what was suggested by their reflection experiments. He explicitly declined to speculate in this paper on what assumptions could be made in order to account for this disparity. [Geiger 1910, p. 500] It would be left to Ernest Rutherford to show in 1911 that this could be explained if the positive charge were concentrated within a radius very small compared to the size of a typical atom.

## 3.3    Competing theories and experimental data

Rutherford had closely examined Thomson's model for β-scattering and concluded that any large-angle deflections of an incident particle due to a continuously distributed sphere of charge [as expressed in equation (10) above] would only be possible if the radius of the sphere were very small compared to the distance over which the interaction took place.  He therefore proposed in his 1911 paper to discuss an atom of "simple structure" that would be capable of producing a large scattering angle, containing a stationary mass of charge ±*Ne* concentrated within a tiny volume.[39]   He planned to show that the scattering due to atomic electrons could essentially be ignored, which would allow him without loss of generality to neutralize the atom with an equal amount of opposite charge spread uniformly across a larger sphere of radius *R*.  In neglecting the internal configuration

---

[38] Typical atomic dimensions could be estimated, for example, with the kinetic theory of gases and an accurate determination of Avogadro's number. This was possible using Einstein's theory of Brownian motion (1905), and the experimental measurements of Siedentopf and Zsigmondy (1903). [Patterson, p. 51]

[39] Rutherford commented in a footnote (but did not elaborate) on how his main deductions were independent of the sign of the central charge. [Rutherford 1911, p. 673] The reason is that the hyperbolic trajectory of an α-particle interacting with a repulsive charge located at the external focus would be equivalent to that caused by an attractive charge located at the internal focus (the α-particle would swing about the atom in a semi-orbit).

of the atomic electrons (because he was only concerned with scattering data), Rutherford was free for the time being to disregard any questions about the dynamic stability of a nuclear atom. [Rutherford 1911, p. 670-1]

He first related the kinetic energy of a typical α-particle to its maximum potential energy inside a neutral atom, in order to find the distance *b* of closest approach to the center:

$$\frac{1}{2}mu^2 = NeE\left(\frac{1}{b} - \frac{3}{2R} + \frac{b^2}{2R^3}\right).$$

**(12)**

[*E* and *m* are the charge and mass of the incident particle, and the subtracted term on the right sets the zero of potential at the surface of the atom.] Substituting $N = 100$ and $u \approx 2\times10^9$ cm/s into this equation (with the assumption $b/R << 1$) he saw that it would be possible for α-particles to penetrate an atom as far as $3.4\times10^{-12}$ centimeters from the center. With the atomic surface at a radius of around $10^{-8}$ cm, he concluded that the influence of the negative charges could be ignored without significant error. Rutherford used the distance $b = 2NeE/mu^2$ as the maximum radius for the nucleus of an atom. [Rutherford 1911, p. 671-2]

He next applied conservation of angular momentum to express the angle of deflection $\phi$ in terms of the *impact parameter p* (the perpendicular distance of the incident trajectory from the center of the atom):

$$\cot\left(\phi/2\right) = \frac{2p}{b}.$$

**(13)**

Since large-angle deflections could only occur for sufficiently small values of *p* (less than $10^{-11}$ cm for both α- and β-particles), he momentarily ignored these to show that the average *small-angle* deflection by the nucleus would be given by

$$\phi_1 = \frac{3\pi}{8}\frac{b}{R},$$

**(14)**

which is three times larger than what was predicted by Thomson for multiple scattering off a uniformly distributed positive sphere. Using the same "random walk" probability argument as Thomson, Rutherford found that for small thicknesses, the chance of scattering a particle into a given angle by a *single* atomic encounter was always greater, and sometimes much greater, than the probability for *compound* scattering to achieve the same result. Single scattering should therefore be the norm rather than the exception in sufficiently thin materials. [Rutherford 1911, p. 678-9]

Rutherford believed the rudimentary nature of the experiments by Geiger and Marsden in 1909 (on the large-angle scattering of α-particles) made them not quite suitable for making real quantitative comparisons, though he did state without proof that the deduced fraction of reflected particles (1 in 8000) would be in rough agreement with his theory if platinum had a central charge of around $100e$. If Bragg's conclusions about the stopping power of a single atom were correct, he argued, the reflection intensity should vary as $A^{3/2}$ (assuming the central charge to be proportional to the atomic weight), and this was also approximately true in those experiments. [Rutherford 1911, p. 680-1; see Table IX]

| Metal. | Atomic weight. | $z$. | $z/A^{3/2}$. |
|---|---|---|---|
| Lead | 207 | 62 | 208 |
| Gold | 197 | 67 | 242 |
| Platinum | 195 | 63 | 232 |
| Tin | 119 | 34 | 226 |
| Silver | 108 | 27 | 241 |
| Copper | 64 | 14.5 | 225 |
| Iron | 56 | 10.2 | 250 |
| Aluminium | 27 | 3.4 | 243 |
| | | | Average 233 |

**Table IX.** Number of scintillations $z$ recorded by Geiger and Marsden for the reflection of α-particles off of metals with different atomic weights $A$, all under similar conditions. The ratio $z/A^{3/2}$ is roughly constant, in agreement with Rutherford's predictions. [From Rutherford 1911, p. 681; compare with Table VII, where Geiger and Marsden computed the ratio $z/A$.]

Rutherford had studied the curves from Geiger's 1910 experimental data (found by plotting the number of scintillations versus scattering angle), and estimated the *most probable* angle of deflection (corresponding to the peak in the curve) to be around 20% less than the *median* scattering angle for each material (the angle inside which half the particles were scattered).[40]  Geiger had measured the most probable scattering angle for a gold foil with thickness $t = 1.7 \times 10^{-4}$ cm to be $1^0$ 40', corresponding to an angle of nearly $2^0$ through which half the particles were scattered, and a value of $N = 97$ for gold.   A similar calculation for a thickness $t = 4.7 \times 10^{-4}$ cm of gold led to a value of $N = 114$.

> "...it follows from these considerations that the magnitude of diffuse reflection of α-particles through more than $90^0$ from gold and the magnitude of the average small angle scattering of a pencil of rays in passing through gold-foil are both explained on the hypothesis of single scattering by supposing the atom of gold has a central charge of about 100*e*." [Rutherford 1911, p. 683]

Rutherford's final comparison between theory and experiment focused on the data presented by Crowther in 1910 on the scattering of β-rays, which had confirmed Thomson's prediction that $\phi / \sqrt{t_m}$ = constant (assuming compound scattering and homogeneous sources).  Rutherford's model for single scattering also predicted this result, so he instead turned to Crowther's calculation of the number of electrons per atom, based on the theory of a uniformly distributed positive sphere. Using Crowther's data and his own model, Rutherford found values for *N* that were much closer to the actual atomic weight of each element. [Table X; see Table V for comparison with Crowther.]

---

[40] Compare this with Crowther's definition of t_m as the thickness of a material required to reduce by half the number of particles scattered inside a fixed angle $\phi$ (defined by a set aperture).

| Element. | Atomic weight. | $\phi/\sqrt{t_m}$ | $N$ |
|---|---|---|---|
| Aluminium | 27 | 4.25 | 22 |
| Copper | 63.2 | 10.0 | 42 |
| Silver | 108 | 15.4 | 78 |
| Platinum | 194 | 29.0 | 138 |

**Table X.** Amount of positive charge $N$ (when multiplied by $e$) for each element, as calculated by Rutherford using his own atomic model and the 1910 data from Crowther on the scattering of homogeneous β-rays. [From Rutherford 1911, p. 685]

"Taking into account the uncertainties involved ... the agreement is sufficiently close to indicate that the same general laws of scattering hold for the α and β particles, notwithstanding the wide differences in the relative velocity and mass of these particles." [Rutherford 1911, p. 685]

Rutherford was conscious of his model lacking full corroboration (because of the poor quality of the available data), and reported that further experiments by Geiger and Marsden were underway to confirm the angular dependence of his theory (their results would not be published until 1913). He also proposed several experiments for addressing some still unresolved issues. Refined large-angle scattering experiments with both α- and β-particles would be better suited for deducing the value of $N$, since errors due to small-angle scattering could be avoided. The sign of the central charge might be determined by comparing predicted absorption rates with the observed absorption of β-rays, which should be more strongly affected by a positive nuclear charge than a negative one.

Although Rutherford was agnostic (at least in this paper) on the actual distribution of atomic electrons, he returned to the question of stability in his final comments, making brief mention of the "Saturnian" model proposed by Hantaro Nagaoka in 1904, but merely to note that Nagaoka had found conditions for the mechanical stability of such an atom, and that Rutherford's conclusions were essentially independent of whether the atom were viewed as a sphere or a disk. [Rutherford 1911, p. 688] Rutherford was giving Nagaoka credit for demonstrating the dynamic stability of a ring of electrons orbiting a massive nucleus (in complete analogy with Maxwell's treatment), but G. A. Schott pointed out shortly after its appearance that Nagaoka's paper contained significant sign errors. [Schott, 1904] There has been no evidence that Nagaoka's work had any direct influence on the development of Rutherford's theory. In a long letter written by Nagaoka to Rutherford in 1910 recounting his recent visit to Europe, and his meeting with Rutherford himself, there was no mention at all of atomic structure. [Badash, p. 55]

There seems to have been little immediate reaction to Rutherford's nuclear atom, and his own reticence may have contributed to this. He did not speak on his model at the first Solvay conference later that year, and in a textbook completed the next year, Rutherford devoted only three pages to the subject of α-scattering. When Niels Bohr met Ernest Rutherford for the first time in November 1911, they did not discuss atomic structure. As Bohr later recalled:

> "At that time, the Rutherford model was not taken seriously. We cannot understand it today, but it was not taken seriously at all. There was no mention of it in any place. [...] I knew how Rutherford looked at the atom, you see, and there was really not very much to talk about." [Quoted in Pais, p. 125]

# 4    The quantum atom of Niels Bohr

## 4.1    Absorption and atomic oscillators

In March 1912, Niels Bohr arrived in Manchester from Cambridge to begin his work with Rutherford, who set him on measuring the absorption of α-particles in aluminum.  Though Bohr came to be known for having had at the time a firm grasp of all the latest developments in physics, both theoretical and experimental, he found that he personally had little enthusiasm for experimental work, and decided to instead concentrate only on theoretical problems.  There was no question in his mind about the significance of the scattering data, but Bohr's first attempt at atomic modeling was more directly influenced by recent theories from Charles Galton Darwin[41] on absorption, and J. J. Thomson on ionization processes.

In a paper published that summer, Darwin had assumed that an α-particle's loss of kinetic energy when traversing matter was primarily due to encounters with atomic electrons (dealing just with absorption, the relatively rare occurrence of large-angle scattering could be ignored).  To simplify the problem, the electrons were taken to be initially at rest, and their electrostatic interactions with each other and with the nucleus were also neglected (i.e., they were treated as free particles).  Further simplifications came about by assuming that the trajectory of the α-particle would be unchanged if it did not penetrate an atom, and that the particle would be effectively shielded from the influence of any electrons in the opposing hemisphere of the atom when it did. [Darwin, p. 901-2]

The absorption profiles and scattering formulas he derived were in approximate agreement with experimental data in certain limits (e.g., high velocities and large atomic cross-sections).[42]   His final equations depended on only two unknown parameters: the number of electrons in an atom $n$, and the atomic

---

[41] Grandson of biologist Charles Robert Darwin.
[42] Darwin had developed his theory using two separate assumptions about the arrangement of the electrons, uniformly distributed either in the volume of a sphere, or on the surface.  His results for both models were roughly the same in most cases, so he couldn't decide between them based on the available data. [Darwin 1912, p. 919]

radius σ.  Taking the latter to be somewhere between $10^{-8}$ and $10^{-9}$ cm for most elements, he was able to conclude:

> "The number of electrons in the atom appears to be intermediate between the atomic weight and its half. [...] In the case of hydrogen it seems probable that the formula for σ does not hold on account of there being only very few electrons in the atom.  If it is regarded as holding then $n = 1$ almost exactly, but σ is very much larger than seems probable." [Darwin 1912, p. 919-20]

Even in 1912, it was still not absolutely clear to physicists that hydrogen contains only a single electron, though there had indeed been experimental evidence for this, one being that hydrogen had never been observed with more than a single unit of positive charge. [Thomson 1912a, p. 672]

Bohr thought many of Darwin's assumptions were objectionable on physical principles alone, so he decided to work on the problem himself using different methods, and continued doing so after returning to Denmark in July 1912 when his fellowship to study abroad had expired.  In an article appearing in January 1913 (communicated by Rutherford), he wrote that the calculation of α-absorption rates would have been fairly straightforward if he had ignored the binding forces on electrons, except that his integrals diverged when calculating the total loss of energy due to the entire material.  Thomson had run into similar difficulties when working out his latest theories on ionization (where he also treated atomic electrons as free particles), but had gotten around this by introducing a cutoff distance on the interaction that was comparable in size to the average distance between atomic electrons.[43]  When contemplating the inclusion of binding forces, Thomson argued that the typical interaction times were orders of magnitude smaller than the vibrational periods of the bound electrons, so that any results obtained by modeling the electrons as free particles would not be appreciably different. [Thomson 1912b, p. 454]

---

[43] Bohr cited Thomson, J. J. (1906). *Conduction of Electricity through Gases*, Cambridge University Press, p. 370-382; see also Thomson, 1912b.

Thomson's choice for a cutoff distance was based on calculations showing that the effects of the more distant electrons had a greater tendency to cancel out than for electrons that were nearer to the scattering trajectory. Bohr felt this criterion would be fine for scattering, but would not be suited for describing the transfer of kinetic energy to bound electrons, because an α-particle's interaction with an individual oscillator would be almost independent of the presence of other electrons. He decided that a different cutoff parameter could be introduced by modeling the atomic electrons as harmonic oscillators (initially at rest), then comparing the interaction times with the induced periods of vibration. The upshot was that the influence of a bound electron on an α-particle's trajectory would be greatest if the induced oscillation period and the interaction time were comparable, and significantly diminished if the interaction time was much longer than the induced period, as with the more distant electrons. [Bohr 1913a, p. 11-12]

Bohr drew an analogy between his theory and the dispersion of light, where the frequency-dependent response of a dispersive medium to different wavelengths of light could be compared with the different electronic oscillation frequencies induced by α-particles passing at various distances.

> "In fact it will be shown, that the information about the number and the frequency of electrons in the atoms, which we get from the theory of dispersion, will enable us to calculate values for the absorption of α-rays for the lightest elements which are in very close agreement with the observed values." [Bohr 1913a, p. 13]

The assumptions used by Bohr in deriving his formula for the rate of α-particle absorption (in particular, that the ratio of the incident velocity to the electronic vibrational frequency be large compared to the effective size of an atom) made his theory well suited for drawing conclusions from experimental data on the refraction and dispersion of light in hydrogen and helium. C. and M. Cuthbertson had recently compiled such data, and had deduced (using Drude's theory) that helium and molecular hydrogen both contained somewhere in the neighborhood of two electrons.[44] Assuming that both contained *exactly* two electrons, Bohr's calculations led to numbers that were very close to the measured absorption rates of both gases (within 10% for hydrogen and 30% for helium). He also found substantial agreement for oxygen and several of the heavier elements.

> "Adopting Prof. Rutherford's theory of the constitution of atoms, it seems that it can be concluded *with great certainty*, from the absorption of α-rays, that a hydrogen atom contains only 1 electron outside the positively charged nucleus, and that a helium atom only contains 2 electrons outside the nucleus." [Bohr 1913a, p. 30-1; emphasis added]

Perhaps just as significant as these important results is the fact that in this paper, Bohr had applied the work of Max Planck in the context of atoms, to estimate the higher vibrational frequencies in molecular oxygen (which were not well known at the time).

> "According to Planck's theory of radiation we further have that the smallest quantity of energy which can be radiated out from an atomic vibrator is equal to $v.k$, where $v$ is the number of vibrations per second and $k = 6.55 \times 10^{-27}$. This quantity must be expected to be equal to, or at least of the same order of magnitude as, the kinetic energy of an electron of velocity just sufficient to excite the radiation." [Bohr 1913a, p. 26-7]

## 4.2    Hypotheses without mechanical foundation

Niels Bohr was not the first to introduce Planck's *quantum of action* into an atomic model. There had been much debate among physicists in the decade prior about the physical interpretation of this fundamental constant, and whether or not

---

[44] Bohr cited C. & M. Cuthbertson, Proc. Roy. Soc. A **83**: p. 166 (1909); and **84**: p. 13 (1910); also Drude, *Ann. d. Phys.* **14:** p. 714 (1904). The Cuthbertson numbers were "somewhat less than 2" electrons for molecular hydrogen, each with vibrational frequency $2.21 \times 10^{16}$; and 2.3 electrons for helium, with vibrational frequency $3.72 \times 10^{16}$. [Bohr 1913a, p. 23-5]

it was a purely thermodynamic quantity.[45]  In 1909, Einstein suspected there might be some connection between $h$ and the fundamental unit of charge, whereas Wien thought that energy quantization was likely due to some kind of universal property of atoms. [Hermann, p. 90]   In 1910, Arthur Haas was inspired to establish a connection between atomic structure and the numerical value for $h$.  He cited a recent German translation of Thomson's book "Electricity and Matter" when he proposed to model a neutral hydrogen atom as a uniform sphere of positive charge (radius $a$), and a single electron (charge $\varepsilon$ and mass $m$) constrained to orbit on the surface of the sphere.[46]  [Haas, p. 262]

Haas noticed that if he divided $h$ into the energy of an electron at the surface of the atom, the result was of the same order of magnitude as the spectral frequencies for hydrogen; he therefore set $h\nu = \varepsilon^2/a$ (where $\nu$ is the frequency of both the emitted radiation and the orbital motion of the electron).  He found the electron's orbital frequency $\nu = \varepsilon/2\pi a\sqrt{am}$ by equating the centripetal force with the electrostatic attraction, and combined this result with his previous ansatz to get an expression for Planck's constant in terms of other measurable quantities: $h = 2\varepsilon\pi\sqrt{am}$ .[47] [Haas, p. 265-6]

"It was in the air to try to use Planck's ideas in connection with such things," Bohr later said of the time. [Quoted in Pais, p. 144] Bohr had also taken note of recent work by J. W. Nicholson, who in 1912 had essentially repeated the calculations performed by Thomson in 1904, but for a hypothetical nuclear-type atom called *protoflourine* (central charge of +5*e*), invented by Nicholson himself to account for certain spectral frequencies in the solar corona. In the process, he also incorporated principles from Planck's theory, by requiring that the energy of an atomic oscillator change only by integral multiples of *h*. Identifying the principle frequency of the protoflourine atom with a spectral line at 3987.1 angstroms, his initial calculations (using the same methods and notation as Thomson) gave the ratio of energy to frequency as:

$$mna^2\omega^2 \cdot 2\pi/\omega = 154.94 \times 10^{-27} \,. \tag{15}$$

[45] Einstein was the first to suggest that Planck's constant was relevant to the description of atoms, and not just Hertzian oscillators. This was acknowledged by Bohr in his July 1913 paper with references to A. Einstein, *Ann. d. Phys.* **17:** 132 (1905); **20:** p. 199 (1906); **22:** p. 180 (1907). [Bohr 1913b, p. 5]
[46] Citing Thomson's work, Haas described hydrogen as the simplest of all the atoms, containing only a single electron ("das als das einfachste aller Atome nur ein einziges Elektron enthaelt"). [Hass, p. 262] With the electron confined to the surface, the force of attraction from the positive sphere would be exactly the same as if all the positive charge were instead located at the center, making his model mathematically equivalent to a nuclear-type atom.
[47] If one were to solve for $a$ in this equation, the result would be the same expression for the ground state radius of hydrogen derived by Bohr in 1913. Haas eventually *did* solve for $a$ in terms of $h$ and $\varepsilon$ in this paper, [Haas, p. 268] but in a roundabout way, and his numerical estimate $a = 1.88 \times 10^{-8}$ cm was larger than Bohr's value of $a = 0.55\times10^{-8}$ cm (by a factor of 3.4). Some of this discrepancy can be attributed to their use of different numerical values for the fundamental constants; some to mistakes in Haas' analysis. In fairness, Haas' main objective was to establish an atomic basis for the physical interpretation of Planck's constant, by deriving an expression for $h$ in terms of $a$ and $\varepsilon$, and not to find $a$ in terms of $h$ and $\varepsilon$.

This equaled nearly 25 multiples of *h*, and Nicholson thought the ratio could be *exactly* 25*h* if the current values for the charge and charge-to-mass ratio of an electron were slightly off. As merely an aside, Nicholson offered an alternative view on how Planck's constant might be related to atoms. His insight was that this ratio of energy to frequency could also be expressed in terms of the total angular momentum of the electrons in their orbit about the nucleus.

> "If, therefore, the constant *h* of Planck has, as Sommerfeld has suggested, an atomic significance, it may mean that the angular momentum of an atom can only rise or fall by discrete amounts when electrons leave or return. It is readily seen that this view presents less difficulty to the mind than the more usual interpretation, which is believed to involve an atomic constitution of energy itself." [Nicholson 1912, p. 679]

In other words, it was more palatable for him to imagine that the angular momentum of a ring of electrons might change by a discrete amount through the loss or addition of a particle, than to think that energy could be anything other than a continuous quantity.

At the end of his first publication from 1913, Bohr had stated that "further information about the constitution of atoms which may be got from experiments on the absorption of α-rays will be discussed in more detail in a later paper." [Bohr 1913a, p. 31] But the paper he promised would *not* be forthcoming. In the months following his return to Denmark, Bohr was in the process of writing up his many ideas on the structure of atoms and, via post, repeatedly turned to Rutherford for guidance and approval. Their correspondence ultimately led to Bohr producing a paper for Rutherford (later known as "The Rutherford Memorandum"[48]) that went unpublished until after Bohr's death, wherein the key theme was stability. Bohr was disturbed by both the mechanical and radiative instability inherent to other atomic models, and told Rutherford he had an idea for resolving these issues, by introducing a new hypothesis, "for which there will be given no attempt at a mechanical foundation (as it seems hopeless)." [Quoted in Pais, p. 137]

In this instance, he was not referring to the quantization of angular momentum, which had already been proposed by Nicholson in 1912. In his seminal paper "On the Constitution of Atoms and Molecules, Part I" (communicated by Rutherford and published in the July 1913 issue of *Philosophical Magazine*), Niels Bohr posited two general hypotheses as the basis of his theory, the first being that atomic electrons move in discrete circular orbits[49] about a massive nucleus, which could be calculated using the "ordinary mechanics." [Bohr 1913b, p. 7] Ignoring radiative instability for the moment, Bohr's initial step was to set the ionization energy $W$ for a single-electron atom equal to the kinetic energy of the electron in its most tightly bound state, which implicitly assumed an inverse-square law of attraction[50] between the electron (mass $m$ and charge $–e$) and the stationary nucleus (charge $E$). The orbital frequency[51] $f$ of the electron in its lowest energy state would then be:

---

[48] A detailed summary is provided in Pais, p. 135-139.

[49] Bohr stated up front that the orbits should be elliptical, but that they could be treated as circular without loss of generality when considering a system containing only a single electron. [Bohr 1913b, p. 4]

[50] Though not explicitly invoked by Bohr, this would follow from the virial theorem, which relates the average kinetic energy of a bound system to its average potential energy. For an inverse-square law of attraction, the kinetic energy is equal to half the negative of the potential energy, making the total energy negative, and equal in magnitude to the kinetic energy. The magnitudes of both quantities are therefore equal to the minimum energy required to liberate the electron from the atom.

[51] Bohr actually used $\omega$ for the orbital frequency, but $f$ is used here to avoid confusion with modern notation for the angular frequency $\omega = 2\pi f$. The orbital frequency $f$ of the electron should not be confused with the frequency $v$ of the emitted photon.

$$f = \frac{\sqrt{2}}{\pi} \frac{W^{3/2}}{eE \cdot m^{1/2}} \,.$$ **(16)**

Bohr's second hypothesis (which could *not* be discussed on a mechanical basis) asserted that electrons transition between energy states by emitting radiation at a single frequency, which was related to the total emitted energy via Planck's constant. He next supposed that the energy radiated during the capture of an electron (initially at rest) by a stationary nucleus would be equal to an integral multiple of *h* times *half* the electron's final orbital frequency; if it settled into the ground state upon capture, the radiated energy would then be exactly equal to the ionization energy.

$$W = \tau h f / 2 \qquad \tau = \text{integer}$$ **(17)**

Substituting this ansatz into (16), Bohr arrived at expressions for the ground state ionization energy, the frequency of rotation, and the orbital radius *a* in terms of fundamental constants:

$$W = \frac{2\pi^2 m e^2 E^2}{\tau^2 h^2} \,, \qquad f = \frac{4\pi^2 m e^2 E^2}{\tau^3 h^3} \,, \qquad 2a = \frac{\tau^2 h^2}{2\pi^2 m e E} \,.$$ **(18)**

He argued that $\tau = 1$ should correspond to the lowest energy state (because this would maximize *W*), then inserted values for *e*, *e/m* and *h* to get numbers that were of the correct order of magnitude for hydrogen:

$$\frac{W}{e} = 13 \text{ volts}, \qquad f = 6.2 \times 10^{15} \text{ s}^{-1}, \qquad 2a = 1.1 \times 10^{-8} \text{ cm}$$

Bohr offered no real justification at this point for why the frequency of the light quanta emitted in the capture of an electron would be equal to *half* its final orbital frequency, except to say that this assumption "...suggests itself, since the frequency of revolution at the beginning of the emission is 0." [Bohr 1913b, p. 5] Without his further elaboration, the only sensible interpretation of this argument is that the radiated frequency should be the *average* of the initial and final frequencies of the electron's orbital motion.  The ultimate justification would be the agreement of his results with experimental data, though Bohr later in this paper appealed to what came to be known as his *correspondence principle*, by showing that for transitions between higher-energy states (ones with *large quantum numbers*), $W = \tau h f / 2$ would logically follow from the requirement that the orbital and spectral frequencies be approximately the same, as would be expected from classical electrodynamics. [Bohr 1913b, p. 13]

He also showed that this ansatz is mathematically equivalent to requiring the angular momentum of the ground state electron to be $h/2\pi$, though his careful language conveyed a reluctance to imply any mechanical interpretation of his theory through the use of macroscopic concepts.  He did comment, however, that Nicholson had recently called attention to a potentially important connection between angular momentum and Planck's constant. [Bohr 1913b, p. 15] Bohr was appreciative of *some* of Nicholson's results, but highly critical of any atomic model where the spectral frequencies were equated with electronic orbital frequencies, which would be inconsistent with the discrete nature of spectral lines, he argued.  If the "ordinary mechanics" were universally valid, a fixed set of spectral frequencies would require the electrons to orbit at constant speeds for finite amounts of time; but this would be impossible, for as soon as the system started radiating, the orbital frequencies of the charges would begin to change immediately. [Bohr 1913b, p. 7]

Bohr had rewritten the laws of physics with his decree that electrons orbit with constant energy, but he truly distinguished himself from those who had come before him by *not* assuming that the spectral frequencies and the orbital frequencies were the same.  A transition between stationary energy states through the emission or absorption of light quanta was "in obvious contrast to the ordinary ideas of electrodynamics, but appears to be necessary in order to account for the experimental facts." [Bohr 1913b, p. 7] Still, Bohr continued to appeal to classical expectations in order to motivate and interpret his theory, which is understandable considering the supreme difficulty of expressing new concepts within the confines of old ideas, as would be required if physicists were to understand and accept his hypotheses.

In his characteristic way of rehashing difficult concepts from a variety of perspectives, Bohr applied electromagnetic theory to again show that the predictions of his model were consistent with classical expectations in the regime of large quantum numbers. Considering a transition between the states $\tau = N$ and $\tau = N - n$ (with $n$ small compared with $N$), he showed that the emission frequency would be approximately $\nu = nf$, which he argued was analogous to the principle Fourier component of the radiation emitted by an electron in an elliptical orbit with frequency $f$.

> "We are thus led to assume that the interpretation of [eq. (17)] is not that the different stationary states correspond to an emission of different numbers of energy-quanta, but that the frequency of the energy emitted during the passing of the system from a state in which no energy is yet radiated out to one of the different stationary states, is equal to different multiples of $f/2$, where $f$ is the frequency of revolution of the electron in the state considered."
> [Bohr 1913b, p.14]

Simply stated, transitions between constant-energy states occurred by the emission or absorption of a *single* photon; the integer values for the quantum number $\tau$ enumerated the energies of the various states, and not the number of photons emitted in a transition.[52]

The most compelling of Bohr's results was his derivation of the Rydberg constant and the hydrogen spectrum in terms of fundamental constants. With the physical meaning of the quantum number $\tau$ now properly interpreted, the energy *difference* between states $\tau = \tau_1$ and $\tau = \tau_2$ for hydrogen could be gotten from (18), by letting $E = e$ and subtracting:

$$W_2 - W_1 = \frac{2\pi^2 m e^4}{h^2}\left(\frac{1}{\tau_2^2} - \frac{1}{\tau_1^2}\right). \tag{19}$$

---

[52] In the first version of this paper, Bohr had entertained (then later abandoned) the possibility that multiple photons were produced in a transition. [Pais 1991, p. 147] If he had given up on this idea, one has to wonder why Bohr didn't just say so in the first place. This was typical of his personal style, to follow the long chain of reasoning that led to his conclusions, rather than presenting a concise summary of final results.

Putting this difference equal to $h\nu$ gave an expression for the frequency of the radiated photon:

$$\nu = \frac{2\pi^2 me^4}{h^3}\left(\frac{1}{\tau_2^2} - \frac{1}{\tau_1^2}\right) \equiv c \cdot R_H\left(\frac{1}{\tau_2^2} - \frac{1}{\tau_1^2}\right). \qquad\qquad \textbf{(20)}$$

Setting $\tau_2 = 2$ and letting $\tau_1$ take on larger integer values would reproduce the formula discovered by Johann Balmer in 1885 for the visible spectrum of hydrogen. $\tau_2 = 3$ corresponded to the infrared spectrum observed by Paschen in 1908, and $\tau_2 = 1$ (or 4, 5...) predicted as yet unobserved spectral lines in the extreme ultraviolet (and extreme infrared).[53] [Bohr 1913b, p. 8-9]

Bohr claimed throughout his life that he had been unaware of Balmer's empirical formula until he was already well along in the development of his theory. He told Rutherford in a January 1913 letter that his work didn't deal at all with the calculation of spectral frequencies. 'Atomic spectra' did not appear on a list Bohr had mailed to Hevesy in February 1913, detailing the "important ideas I have used as the foundation of my calculations." The version of the paper he was working on in March 1913 contained a complete derivation of the formula. In April, the final version was finished. [Pais, p. 144]

> "I think I discussed [the paper] with someone ... that was Professor Hansen ... I just told him what I had, and he said 'But how does it do with the spectral formulae?' And I said I would look it up, and so on. [...] I didn't know anything about the spectral formulae. Then I looked it up in the book of Stark ... other people knew about it but I discovered it for myself.'" [Quoted in Pais, p. 144]

By combining Rutherford's nuclear model with deductions based on two almost untenable hypotheses, Bohr succeeded where others had failed – spectral frequencies that were in complete agreement with experimental observation. The Balmer formula was known for decades, but had hitherto no theoretical explanation.

---

[53] In 1914, T. Lyman observed two spectral lines for hydrogen that were in the ultraviolet, but did not mention Bohr's theory in his paper. Three new lines in the Paschen series were discovered by F. Brackett in 1922, as well as two lines corresponding to $\tau_2 = 4$. In 1924, A. Pfund measured a single spectral line consistent with $\tau_2 = 5$. [Kragh, p. 69]

Paraphrasing Imre Lakatos, this is a classic example from the history of science regarding the novel prediction of phenomena, rather than the prediction of novel phenomena.

## 4.3    Ionized helium and lithium

The reaction in the scientific community to the Bohr model, which was admittedly an *ad hoc* mixture of both classical and quantum rules, was itself a mixture of skepticism, praise, derision and silence.[54]   Before the paper had even been published, Ernest Rutherford (though delighted with the overall quality of the work) wrote to convey some of his initial misgivings.

> "There appears to me one grave difficulty in your hypothesis, which I have no doubt you fully realize, namely how does an electron decide what frequency it is going to vibrate at when it passes from one stationary state to the other? It seems to me that you have to assume that the electron knows beforehand where it is going to stop." [Quoted in Pais, p. 153]

Rutherford was the first of many to express discomfort with the acausality of "quantum jumps," which continued as a long-standing objection to quantum mechanics.  Albert Einstein, who had been the first to point out the importance of incorporating Planck's constant into atomic physics, ironically became one of the greatest critics of quantum theory, precisely for its clash with his deterministic views of nature.  But Einstein was extremely impressed with Bohr's accomplishment at the time, particularly when he first heard of its unprecedented success in describing the spectrum of ionized helium: "This is an enormous achievement.  The theory of Bohr must then be right." [Pais, p. 154]

Nearly two decades prior, E. Pickering had observed a series of spectral lines from the star $\zeta$ Puppis that converged to the same limiting frequency as hydrogen, and could be described by Balmer's formula using half-integers instead of integers, which led scientists to erroneously take them for a second spectrum of hydrogen.[55] If the hydrogen emitting this "second set" of Balmer lines could only exist in the low pressures of outer space, that might explain why the lines had never been observed in the laboratory. That is, except by A. Fowler in 1912, who had detected several Pickering lines in a laboratory mixture of hydrogen and helium gas. [Bohr 1913b, p. 10] Despite being unable to reproduce the lines with a pure hydrogen sample, Fowler inexplicably maintained they could not be due to helium. [Kragh, p. 69] Bohr's theory also addressed this phenomenon, and could even explain why experimentalists had only seen 12 of the Balmer lines under laboratory conditions, whereas 33 had been identified astronomically. The orbital radii of his atoms increased by the square of the quantum number $\tau$, [eq. (18)] and the much greater size of the atoms sufficiently excited to be in the $\tau = 33$ state (compared with $\tau = 12$) was such that they could only exist in regions of extremely low density (such as nebular clouds). For the intensity of the lines to be observable, he argued, the gas would also have to exist in large quantities, meaning these emission lines would likely never be observed in terrestrial vacuum chambers. [Bohr 1913b, p. 9-10]

Furthermore, Pickering lines could be attributed to ionized helium by simply recasting the spectral formula in accordance with his theory:

$$ \nu = \frac{2\pi^2 m e^4}{h^3} \left( \frac{1}{\left(\tau_2/2\right)^2} - \frac{1}{\left(\tau_1/2\right)^2} \right) = \frac{2\pi^2 m e^2 (2e)^2}{h^3} \left( \frac{1}{\left(\tau_2\right)^2} - \frac{1}{\left(\tau_1\right)^2} \right). \qquad \textbf{(21)} $$

The spectrum could thus be ascribed to an atom with a central charge of $E = +2e$ and a single orbiting electron. A colleague in the Manchester group had been studying Pickering lines in the laboratory while Bohr was developing his theory, and soon reported their detection using a pure sample of helium in a letter to *Nature*,[56] making what is likely the earliest reference to the Bohr model in the scientific literature. [Kragh, p. 71]

---

[55] Every *second* line in the series corresponded to a normal line in the Balmer series, hence the ability to use half-integers in the regular formula.
[56] E. Evans, The spectra of helium and hydrogen, *Nature* **92:** 5 (4 September 1913).

Fowler, unconvinced, pointed out just weeks later in his own letter to *Nature*[57] that there were noticeable discrepancies between the observed Pickering lines and the theoretical values – experimentally, the ratio of $R_{He}$ to $R_H$ should be replaced by 4.0016, and not a simple factor of four.  Bohr, undeterred, shot back in late October,[58] after confirming his suspicion that the error was due to not having accounted for the finite mass of the nucleus.  He replaced the electron mass with the reduced mass of the two-body system $\mu = mM/(m+M)$ to get $R_{He}/R_H = 4.00163$, in agreement with experimental data to five significant figures.  Fowler conceded his defeat immediately.[59]

Bohr dealt with a similar issue in the second of his 1913 trilogy of papers on the structure of atoms and molecules, in this case regarding lines reported by Nicholson in astronomical systems that also exhibited the Pickering spectrum, but which followed a different Balmer-like formula:

$$v = R_H \left( \frac{1}{4} - \frac{1}{\left(m \pm 1/3\right)^2} \right). \tag{22}$$

For the same reasons as before, Nicholson attributed these lines to hydrogen.  Bohr perceived that the lines could be derived with his theory if he re-wrote the frequency formula as

$$v = \frac{2\pi^2 m e^4}{h^3} \left( \frac{1}{\left(\tau_2/3\right)^2} - \frac{1}{\left(\tau_1/3\right)^2} \right), \tag{23}$$

and made the judicious choice of $\tau_2 = 6$.  The spectral formula thus corresponded to an atom with a single electron and a central charge of $+3e$ (doubly-ionized lithium).  The three lines in question were found by letting $\tau_1 = 10$, 13, and 14. [Bohr 1913c, p. 490-1]

Nicholson, whose mechanical nuclear atom was the only serious rival to the quantum atom at the time, was not impressed. In fact, he was to be Bohr's most vocal critic in those first years. His objections were many, among them having to do with mechanical instability (concentric rings would not be stable) and valency (he thought lithium should be an inert gas). Bohr initially intended to reply to Nicholson's criticisms (he even began unfinished letters to *Nature* and *Philosophical Magazine* in response), but apparently decided in the end that it was not worth the effort, and just ignored him. [Kragh, p. 112-6]

The most notable silence came from J. J. Thomson, who gave public lectures on atomic structure in 1914, but made no mention of Bohr and his model. [Pais, p. 153] Thomson remained convinced that mechanical models could be used just as well to explain quantum phenomena, such as the photoelectric effect, and presented in September 1913 a strange atom devised to do just that. It bore little resemblance to his previous work: a mixture of corpuscles in static equilibrium with hydrogen ions and α-particles, where an attractive inverse-square law inside the atom was supplemented by a selectively-acting repulsive inverse-cube force. [Kragh, p. 108] Thomson never gave up on the use of mechanical analogies in atomic physics, but the success of the quantum atom eventually made his models all but irrelevant to the advancement of knowledge. Still, despite the terseness of his sole reference to Bohr in his autobiographical "Recollections and Reflections" (written at the age of 80), Thomson was ultimately and undeniably complimentary:

> "At the end of 1913 Niels Bohr published the first of a series of researches on spectra, which it is not too much to say have in some departments of spectroscopy changed chaos into order, and which were, I think, the most valuable contribution which the quantum theory has ever made to physical science." [Thomson 1937, p. 425]